\documentclass{aa}  
\usepackage{amssymb}
\usepackage{graphicx}
\usepackage[authoryear]{natbib}
\usepackage[figuresright]{rotating}

\usepackage{txfonts}
\newcommand{\hei}{\ion{He}{i} 2.058 $\mu m$}
\newcommand{\heid}{\ion{He}{i} 2.112 $\mu m$}
\newcommand{\heie}{\ion{He}{i} 2.184 $\mu m$}
\newcommand{\heif}{\ion{He}{i} 2.150 $\mu m$}
\newcommand{\heig}{\ion{He}{i} 2.161 $\mu m$}
\newcommand{\heii}{\ion{He}{ii} 2.189 $\mu m$}
\newcommand{\heiib}{\ion{He}{ii} 2.037 $\mu m$}
\newcommand{\heiic}{\ion{He}{ii} 2.346 $\mu m$}
\newcommand{\brg}{Br$_{\gamma}$}
\newcommand{\niiia}{\ion{N}{iii} 2.115 $\mu m$}
\newcommand{\niii}{\ion{N}{iii} 2.247, 2.251 $\mu m$}
\newcommand{\mgii}{\ion{Mg}{ii} 2.138 $\mu m$}
\newcommand{\mum}{\ifmmode \mu \rm{m} \else $\mu \rm{m}$\fi}

\newcommand{\teff}{\ifmmode T_{\rm eff} \else T$_{\mathrm{eff}}$\fi}
\newcommand{\logg}{\ifmmode \log g \else $\log g$\fi}
\newcommand{\lL}{\ifmmode \log \frac{L}{L_{\odot}} \else $\log \frac{L}{L_{\odot}}$\fi}
\newcommand{\mdot}{$\dot{M}$}
\newcommand{\myr}{M$_{\odot}$ yr$^{-1}$}
\newcommand{\vsini}{$V$ sin$i$}
\newcommand{\vinf}{$v_{\infty}$}
\newcommand{\vturb}{v$_{\rm turb}$}

\newcommand{\kms}{km s$^{-1}$}
\newcommand{\msun}{\ifmmode M_{\odot} \else M$_{\odot}$\fi}
\newcommand{\zsun}{\ifmmode Z_{\odot} \else Z$_{\odot}$\fi}
\newcommand{\lsun}{\ifmmode L_{\odot} \else L$_{\odot}$\fi}
\newcommand{\rsun}{\ifmmode R_{\odot} \else R$_{\odot}$\fi}
\newcommand{\qh}{\ifmmode Q_{\rm H} \else $Q_{\rm H}$\fi}
\newcommand{\qhei}{\ifmmode Q_{\ion{He}{i}} \else $Q_{\ion{He}{i}}$\fi}

\begin{document}
   \title{Stellar and wind properties of massive stars in the central parsec of the Galaxy}

   \subtitle{}

   \author{F. Martins\inst{1}
          \and
          R. Genzel\inst{1,2}
	  \and
	  D.J. Hillier\inst{3}
	  \and
	  F. Eisenhauer\inst{1}
	  \and
	  T. Paumard\inst{1,4}
	  \and
	  S. Gillessen\inst{1}
	  \and
	  T. Ott\inst{1}
	  \and
	  S. Trippe\inst{1}
          }

   \offprints{F. Martins}

   \institute{Max-Planck Instit$\ddot{\rm u}$t f$\ddot{\rm u}$r extraterrestrische Physik, Postfach-1312, D-85741, Garching, Germany \\
              \email{martins@mpe.mpg.de}
         \and
             Department of Physics, University of California, Berkeley, CA 94720, USA\\
         \and
	     Department of Physics and Astronomy, University of Pittsburgh, 3941 O'Hara St., Pittsburgh, PA 15260, USA\\
	 \and
	     LESIA, Observatoire de Paris, 5 Place Jules Janssen, F-92195, Meudon, France\\
             }

   \date{Received 3 November 2006 / Accepted 1 March 2007}

\titlerunning{Massive stars in the Galactic Center}
\authorrunning{F. Martins et al.}


\abstract
   {How star formation proceeds in the Galactic Center is a debated question. Addressing this question will help us understand the origin of the cluster of massive stars near the supermassive black hole, and more generally starburst phenomena in galactic nuclei. In that context, it is crucial to know the properties of young massive stars in the central parsec of the Galaxy.}
   {The main goal of this study is to derive the stellar and wind properties of the massive stars orbiting the supermassive black hole SgrA$^{\star}$ in two counter-rotating disks.}
   {We use non-LTE atmosphere models including winds and line-blanketing to reproduce H and K band spectra of these stars obtained with SINFONI on the ESO/VLT.}
   {The GC massive stars appear to be relatively similar to other Galactic stars. The currently known population of massive stars emit a total $6.0 \times 10^{50}$ s$^{-1}$ (resp. $2.3 \times 10^{49}$ s$^{-1}$) H (resp. \ion{He}{i}) ionising photons. This is sufficient to produce the observed nebular emission and implies that, in contrast to previous claims, no peculiar stellar evolution is required in the Galactic Center. We find that most of the Ofpe/WN9 stars are less chemically evolved than initially thought. The properties of several WN8 stars are given, as well as two WN/C stars confirmed quantitatively to be stars in transition between the WN and WC phase. We propose the sequence (Ofpe/WN9 $\rightleftharpoons$ LBV) $\rightarrow$ WN8 $\rightarrow$ WN/C for most of the observed GC stars. Quantitative comparison with stellar evolutionary tracks including rotation favour high mass loss rates in the Wolf-Rayet phase in these models. In the OB phase, these tracks nicely reproduce the average properties of bright supergiants in the Galactic Center.}
   {}

\keywords{Stars: early type - Stars: Wolf-Rayet - Stars: atmospheres -
Stars: fundamental parameters - Stars: winds, outflows - Galaxy:
center}

\maketitle


\section{Introduction}
\label{intro}

The center of our Galaxy is a unique environment to study massive
stars. It harbors three of the most massive clusters of the Galaxy --
the Arches, Quintuplet and central clusters. Heavily extincted and
only accessible at infrared (and longer) wavelengths or in X-rays,
each of these clusters has a population of more than a hundred massive
stars. Even more interesting is the difference in their ages: 2.5, 4
and 6 Myrs for the Arches, the Quintuplet and the central cluster
respectively \citep{figer99,figer02,pgm06}. Such a spread implies the
presence of different types of massive objects, naturally sampling
stellar evolution in the upper HR diagram. The youth of these
clusters, together with their total mass in excess of $10^{4}$ \msun,
partly explains the large number of massive stars in the Galactic
Center. But another reason may be the top-heavy mass function:
\citet{stolte02} for the Arches and \citet{pgm06} for the central
cluster have shown that the slope $\Gamma$ of the present-day mass
function was shallower than the standard Salpeter value ($-0.8$
instead of $-2.35$). This may be due to mass segregation or to a true
feature of the \textit{initial} mass function. In that case the
Galactic Center could be a peculiar environment for the formation of
massive stars.

As a peculiar environment, the central cluster is especially
interesting since it harbors the supermassive black hole
SgrA$^{\star}$. The first stars ever observed in this region were the
so-called ``AF'' and ``IRS16'' stars \citep{forrest87,ahh90}. Their
K-band spectrum showed strong emission lines -- in particular the
\hei\ feature -- and they were immediately classified as Ofpe/WN9, a
class of evolved massive stars. Further observations by
\citet{krabbe91} revealed a few additional \ion{He}{i} emission line
stars. They argued that this population most probably resulted from a
burst of star formation a few Myrs ago \citep{krabbe95}. Detailed
spectroscopic analysis with atmosphere models by \citet{paco94,paco97}
established that these objects were post-main sequence massive
stars. Their UV flux was able to ionise the ISM and produce the
nebular \brg\ emission, but was far too soft to explain the nebular
\ion{He}{i} emission. The presence of a population of hotter objects not
accessible to observations was thus inferred.

This population was unraveled in the last years. \citet{paumard01}
found additional stars with broader lines than the initial ``He~{\sc
i}'' stars and typical of Wolf-Rayet stars. \citet{genzel03} deduced
the presence of even more massive stars from the absence of CO
absorption bands. The advent of adaptive-optics assisted integral
field spectroscopy lead to a new breakthrough: \citet{pgm06}
spectroscopically identified nearly a hundred massive stars, including
various types of Wolf-Rayet stars, O and B supergiants, and even
dwarfs \citep[see also][]{horrobin04,paumard04}. Previously,
\citet{ghez03} and \citet{frank05} had shown that the group of stars
located in the central arcsecond of the Galaxy and orbiting very close
to the black hole (the so-called 'S-stars') was composed of early to
late B dwarfs.

This latter group of stars is at the heart of an issue usually
referred to as the ``paradox of youth'' of the Galactic center: how
could star formation (traced by the presence of young massive stars)
have happened so close to the black hole, where the tidal forces
should prevent any molecular cloud from collapsing \citep{morris93}?
Two scenarios are invoked to solve this problem. In the ``in-situ
formation'' scenario, a disk forms around the supermassive black hole
with a density large enough to be self-gravitating so that tidal
forces do not perturb the collapse of molecular material
\citep[e.g.][]{lb03,nc05}. In the alternate ``in-spiraling cluster''
scenario, young massive stars are born in a dense cluster several tens
of parsec away from SgrA$^{\star}$ and its hostile environment, and
are subsequently brought to the central region by spiral-in of the
parent cluster due to dynamical friction
\citep[e.g.][]{gerhard01,km03}. The former picture (star formation in
a self-gravitating disk) appears more attractive since it accounts for
a larger number of observational facts \citep{pgm06}, but the question
is not completely settled.

In that context, it is especially important to get a better knowledge
of the physical properties of the massive stars in the central parsec
of the Galaxy. In parallel to the development of powerful infrared
observational techniques, reliable atmosphere models have become
available in recent years. This is due to a huge effort from different
groups to include thousands of metallic lines in such models
(``line-blanketing''). The use of such models has quantitatively
changed our knowledge of the stellar and wind properties of massive
stars. To name a few, their effective temperatures are lower
\citep{msh02,paul02,repolust04} and their winds are highly clumped
\citep{hil03,jc05,fullerton06}. In addition to these significant
developments, special attention was given to the infrared
range. Several studies have been undertaken to compare the parameters
derived from IR diagnostics with those obtained from classical optical
features \citep{bc99,repolust05}. It appears that the differences are
usually small, and certainly within the uncertainties on the derived
parameters. Then, both new observational data and better atmosphere
models are now available. A quantitative investigation of the stellar
and wind properties of massive stars in the central parsec of the
Galaxy, in the context of the ``paradox of youth'', is thus possible.

We present such a study in this paper. Sect.\ \ref{observ} describes
the observational data. In Sect.\ \ref{modeling}, we present our
atmosphere models and explain our method to derive stellar and wind
properties. This is followed by a detailed analysis of individual
stars (Sect.\ \ref{results}). We then discuss the results with special
emphasis on ionising fluxes (Sect.\ \ref{dis_ion}), stellar evolution
(Sect.\ \ref{dis_evol}), metallicity (Sect.\ \ref{dis_z}), stellar
winds (Sect.\ \ref{dis_wind}) and chemical composition of OB stars
(Sect.\ \ref{dis_ob}). We finally summarize our findings in Sect.\
\ref{conc}.


\section{Observations}
\label{observ}

The observations analysed here were conducted with the integral field
spectrograph SPIFFI/SINFONI on the ESO/VLT Yepun 8 meter telescope
\citep{sinfoni,spiffi,bonnet04} as part of the MPE-Garching GTO
program ``Galactic Center''. A first mosaic of data cubes was obtained
on Apr. 8$^{\rm th}$ 2003 using the 250 mas scale in the K band mode,
allowing a resolution of $\sim$ 4000. A second mosaic was observed in
H+K band on the same scale on Apr. 9$^{\rm th}$ 2003. These
observations were conducted in seeing limited mode since at that time
the AO system MACAO was not yet coupled to SPIFFI. A new SINFONI
mosaic was obtained on Aug. 18-19$^{\rm th}$ 2004 in K band with the
100 mas scale in adaptive optics mode. Finally, several fields of the
region $\sim$ 15 '' north of SgrA$^{\star}$ were observed on
Mar. 16-17$^{\rm th}$ 2005. Additional information on all these data
cubes can be found in \citet{pgm06}. Two new fields were obtained in
the configuration H+K/0.1 mas scale on Apr. 20$^{th}$ 2006 and
Aug. 16$^{th}$ 2006.

\begin{figure}
\includegraphics[width=9cm]{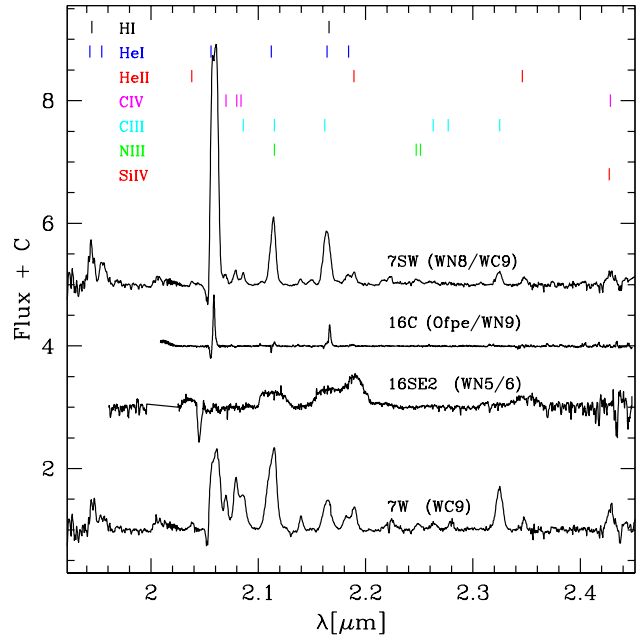}
\caption{SINFONI K-band spectra of four post main sequence massive
stars observed in the GC. The main lines are marked. Note that all
lines are not present simultaneously in one spectrum: depending on the
stellar and wind parameters (in particular \teff\ and abundances),
only a subset of the marked lines is observed in a given
star. \label{line_id}}
\end{figure}

To extract the spectra out of these cubes, we defined ``source''
pixels showing spectral signatures of the massive stars identified by
\citet{pgm06} from which we removed neighboring ``continuum'' sources
to correct for the local background. In the definition of the
``continuum'' pixels in crowded regions, we paid special attention not
to include pixels contaminated by neighboring stellar sources. We used
several combinations of source-continuum pixels to check the
reliability of the extracted spectrum. We found that as long as
contaminating sources are not included, the spectra can safely be
extracted. Fig.\ \ref{line_id} shows typical spectra of various types
of stars analysed here together with line identification.

\section{Modeling}
\label{modeling}

\subsection{Atmosphere models}
\label{atm_models}

In order to derive the stellar and wind properties of the massive
stars in the central parsec of the Galaxy, we have used state of the
art atmosphere models computed with the code CMFGEN \citep{hm98}. Such
models include the main ingredients necessary to produce realistic
atmospheric structures and emergent spectra, namely a non-LTE
treatment, winds and line-blanketing. The latter has been included
only recently since in combination with the two former ingredients it
leads to complicated and numerically demanding simulations. But the
effects of line-blanketing are qualitatively and quantitatively very
important. Most stellar parameters have to be revised when determined
with line-blanketed models due to the strong modification of the
radiative transfer caused by additional opacities from metals.

CMFGEN computations proceed in two steps: first the atmospheric
structure and radiative field are computed in an iterative process;
second, a formal solution of the radiative transfer equation with
opacities given by the first step is determined, leading to the
detailed emergent spectrum. Details about the code CMFGEN are
discussed by \citet{hm98} and here we only recall the main
characteristics:

\begin{itemize}

\item \textit{non-LTE treatment:} populations of individual energy
levels are computed through the resolution of statistical equilibrium
equations including radiative and collisional processes. Non-LTE is
especially important for infrared studies since in this range,
stimulated radiative processes are greatly enhanced compared to
shorter wavelengths \citep{annique04} and thus have a strong impact on
level populations.

\item \textit{winds:} a spherical geometry is adopted to fully take
into account the atmospheric extension due to winds. Velocity
gradients in the accelerating wind are also included in the radiative
transfer problem. An important point to be noted here is the fact that
a velocity law (equivalent to a density law through the equation of
mass conservation) has to be adopted since radiative acceleration is
not used to compute the hydrostatic structure of the atmosphere. In
practice, two approaches are used: for OB stars, a photospheric
structure computed with another atmosphere code \citep[usually TLUSTY,
see][]{hl95} is smoothly connected to a so called ``$\beta$ velocity
law ($v = v_{\infty}(1-\frac{R_{\star}}{r})^{\beta}$) where \vinf\ is
the terminal velocity and $R_{\star}$ the stellar radius; for stars
with denser winds, the atmosphere is usually optically thick so that
the inner velocity structure is less crucial and a law of the form

\begin{equation}
v = v_{0} + \frac{(v_{\infty} - v_{0})(1-\frac{R_{\star}}{r})^{\beta}}{1+\frac{v_{0}}{v_{\rm core}}e^{\frac{R_{\star}-r}{h_{\rm eff}}}}
\end{equation}
\label{eq:vellaw}

\noindent is adopted ($v_{\rm core}$ being the velocity at the bottom of
the atmosphere, $v_{0}$ the velocity at the expected photospheric
velocity and $h_{\rm eff}$ the density scale height of the
photosphere). $\beta$ can in principle be derived from the shape of
spectral lines \citep{puls96,n81} but here, it is only possible in a
few cases. Hence, for most of the analysis, we simply adopt the
standard value 1.0 \citep{paco94,paulwc9}.

CMFGEN also allows the treatment of non homogeneous atmosphere
through the adoption of a clumping law of the form

\begin{equation}
f = f_{\infty} + (1-f_{\infty})e^{-\frac{v}{v_{\rm init}}}
\end{equation}
\label{eq:clumping}

\noindent where $f_{\infty}$ is the value of $f$ at the top of the
atmosphere and $v_{\rm init}$ is the velocity at which clumping
appears. We adopted a typical value of $f_{\infty} = 0.1$ for the
remaining stars, unless explicitly indicated.

\item \textit{line-blanketing:} CMFGEN includes a direct treatment of
metals. The main approximation\footnote{Other
approximations concern the treatment of line profiles, redistribution
functions and microturbulence.} (which can be easily dropped provided
the computational resources are available) is the grouping of levels of
similar energies in ``super-levels'' as initially proposed by
\citet{and91}. Through the resolution of the statistical equilibrium
equations, level population of metals are computed as for H and He. In
the models presented here, C, N, O, Ne, Na, Mg, Al, Si, S, Ar, Ca, Fe,
Ni are included for Wolf-Rayet stars and related objects, while C, N,
O, Si, S and Fe are used for OB stars. This is due to the lower
density winds of the latter which require a better spatial
sampling. In that case, minor metals have to be dropped to keep the
size of the models reasonable. Note that no super-levels were
used for H, \ion{He}{i} and \ion{He}{ii} so that lines from these
elements do not suffer from this approximation. In WC stars, we also
did not use super-levels for \ion{C}{iii} and \ion{C}{iv}. Finally, we
tested the effect of super-levels on the \niii\ lines in WN stars and
found it was negligible.  In the following, we refer to \citet{gs98}
for the solar abundances.

\item \textit{temperature structure:} the temperature structure in the
atmosphere is set under the condition of radiative equilibrium.

\item \textit{microturbulent velocity:} a microturbulent velocity
(\vturb) must be provided in the computation of both the atmospheric
structure and the detailed emergent spectrum. In the former part of
the simulation, \vturb\ is independent of the position in the
atmosphere and we usually choose typical values of 20 \kms\ for OB
stars and 50 \kms\ for Wolf-Rayet stars. For the computation of the
detailed emergent spectrum, we adopt \vturb\ = 10 \kms. Note
that we have run test models with larger values of \vturb\ for the
computation of the emergent spectrum, and found no difference since
the lines are mainly shaped by the wind terminal velocity, much
larger than \vturb.

\end{itemize}

\subsection{Method}
\label{method}

The main diagnostics we used for the analysis were: 1) K band
photometry and 2) normalized spectra. The number of diagnostics is
relatively limited, which sometimes results in degeneracies in the
derived parameters, especially when \teff\ cannot be accurately
constrained. The procedure to derive the stellar and wind parameters
is as follows:

\begin{itemize}

\item \textit{Effective temperature:} we relied on the ratio of lines
from successive ionisation states of a given element: He for OB and WN
stars, He and C for WN/C and WC stars. In practice, the following
lines were used: \ion{He}{i} 2.112 \mum, \ion{He}{i} 2.184 \mum,
\ion{He}{ii} 2.037 \mum, \ion{He}{ii} 2.189 \mum, \ion{He}{ii} 2.346
\mum, \ion{C}{iv} 2.070 \mum, C~{\sc iv} 2.079 \mum, \ion{C}{iv} 2.084
\mum, \ion{C}{iii} 2.325 \mum. The \ion{He}{i} line at 2.058 \mum\ has
been shown to greatly depend on basically \textit{any} parameter
\citep{paco94} and, although it is frequently the strongest observed
line, it is not used to derive \teff. Unfortunately, two successive
ionisation ratios of the same element are not always present,
especially for the latest OB and Wolf-Rayet stars. In that case, one
usually sees only \ion{He}{i} lines so that only an upper limit on
\teff\ can be derived.

We also define here a temperature $T_{\star}$ such that

\begin{equation}
L = 4 \pi R_{\star}^{2} T_{\star}^{4} = 4 \pi R_{2/3}^{2} \teff^{4}
\end{equation}
\label{eq:Tstar}

\noindent $R_{\star}$ is the radius at which the Rosseland optical
depth $\tau_{\rm Rosseland}$ is equal to 20 ($R_{2/3}$ corresponding
to the radius where the $\tau_{\rm Rosseland} = 2/3$). This definition
is useful in the case of evolved massive stars since it is not
directly affected by the stellar wind, and since it allows a better
comparison of stellar parameters to evolutionary models (see Sect.\
\ref{dis_z_hei}).\\

\item \textit{Luminosity:} the main constraint on the luminosity comes
from the absolute K magnitude (M$_{\rm K}$) which is derived from the
observed m$_{\rm K}$, the distance to the Galactic Center \citep[7.62
kpc according to][]{frank05}, and the extinction taken from the recent
work of \citet{schoedel07} (see Table \ref{tab_obs}). Knowing the
distance of the sources is a great advantage over many Galactic
studies of massive stars. The main source of uncertainty in the
luminosity is the extinction which is known to vary in the central
parsec.  A given value of M$_{\rm K}$ is obtained for a given flux in
the K band which depends on 1) the luminosity and 2) on the wind
density. Indeed, free-free emission in the atmosphere can produce an
excess of emission mimicking a larger luminosity \citep{lc99}. Hence,
luminosity is derived in combination with wind parameters (\mdot\ and
\vinf).\\

\item \textit{Mass loss rate:} \mdot\ is derived from the strength of
emission lines formed by recombination. Their intensity depends on the
wind density but also on the ionisation of the atmosphere (which is
also partly controlled by the density) and on the abundances. The mass
loss rate is thus derived in combination with effective temperature
and abundances. \\

\item \textit{Terminal velocity:} In stars showing blueshifted
absorption profiles and P-Cygni lines, \vinf\ was estimated from the
velocity shift of the bluest part of the absorption trough with
respect to the rest-frame wavelength. For Wolf-Rayet stars with broad
emission lines, half of the line width was chosen as typical of
the terminal velocity. Values were then refined according to the
quality of the fit. \\

\item \textit{Abundances:} The relative H to He abundances can be
derived from the intensity of various lines from these elements. As
mentioned previously, such a determination cannot be separated from
the estimate of \mdot\ and \teff\ so that a simultaneous determination
of these parameters is done.  In stars showing C and N lines (C~{\sc
iv} 2.070 \mum,\ion{C}{iv} 2.079 \mum, \ion{C}{iv} 2.084 \mum,
\ion{C}{iii} 2.325 \mum, \ion{N}{iii} 2.247 \mum, \ion{N}{iii} 2.251
\mum), constraints on carbon and nitrogen abundances can also be
given. Note that we do not use the feature at 2.115 \mum\ for the N
abundance determination: it is unclear whether \ion{N}{iii} is the
only contributor \citep{geballe06} and, in addition, the transition
oscillator strength is uncertain.\\

\item \textit{Mass:} The determination of masses for hot stars is a
challenge. Here, we have used the $M - L$ relation of \citet{hl96} to
estimate the present mass of H free Wolf-Rayet stars. For other stars,
we did not try to derive masses, the uncertainties being too large.\\

\end{itemize}

The uncertainties in the derived parameters are the following: $\pm$
3000 K for \teff ($\pm$ 6000 K for the Ofpe/WN9 stars), $\pm$ 0.2 in
\lL, 0.2 dex for \mdot, 30 \% for the abundances. They are
\textit{not} statistical errors (the estimate of such errors would
imply the computation of a huge number of models to sample the
parameter space around the best fit solution). Instead, they reflect
the range of values leading to acceptable fits of the observed
spectra. Our errors also do not include any systematic
contribution due to uncertainties in atomic data such as collisional
and dielectronic recombination cross-sections. This has to be kept in
mind when considering the values we quote.

\begin{table*}
\begin{center}
\caption{Observational properties of the stars analyzed in this paper. Spectral types are from \citet{pgm06}. Values of extinction are taken from \citet{schoedel07}. A distance of 7.62 kpc is adopted \citep{frank05}. \label{tab_obs}}
\begin{tabular}{llrrrl}
\hline\hline
Star    & ST        & m$_{\rm K}$  & A$_{\rm K}$ & M$_{\rm K}$  & alternative name \\
\hline                						  
34W     &  Ofpe/WN9 & 12.5$\pm$0.1 & 3.0$\pm$0.2 & -6.0$\pm$0.3 & GCIRS 34W        \\
16NW    &  Ofpe/WN9 & 10.0$\pm$0.1 & 2.3$\pm$0.1 & -6.7$\pm$0.2 & GCIRS 16NW       \\
16C     &  Ofpe/WN9 &  9.7$\pm$0.1 & 2.3$\pm$0.1 & -7.0$\pm$0.2 & GCIRS 16C        \\ 
33E     &  Ofpe/WN9 & 10.1$\pm$0.1 & 2.5$\pm$0.1 & -6.8$\pm$0.2 & GCIRS 33E        \\
AF      &  Ofpe/WN9 & 10.8$\pm$0.1 & 2.2$\pm$0.1 & -5.8$\pm$0.2 & NAME AF STAR     \\ 
15NE    &  WN8      & 11.8$\pm$0.1 & 2.0$\pm$0.2 & -4.6$\pm$0.3 & GCIRS15 NE       \\
AFNW    &  WN8      & 11.7$\pm$0.1 & 2.3$\pm$0.2 & -5.0$\pm$0.3 & NAME AF NW       \\
9W      &  WN8      & 12.1$\pm$0.1 & 2.7$\pm$0.2 & -5.0$\pm$0.3 & GCIRS 9W         \\
7E2     &  WN8      & 12.9$\pm$0.1 & 2.6$\pm$0.2 & -4.1$\pm$0.3 & GCIRS 7E2        \\
13E2    &  WN8      & 10.8$\pm$0.1 & 2.0$\pm$0.3 & -6.4$\pm$0.4 & GCIRS 13E2       \\
7SW     &  WN8/WC9  & 12.0$\pm$0.1 & 2.8$\pm$0.2 & -5.2$\pm$0.3 & $-$              \\
15SW    &  WN8/WC9  & 12.0$\pm$0.1 & 2.0$\pm$0.2 & -4.4$\pm$0.3 & GCIRS 15SW       \\
AFNWNW  &  WN7      & 12.6$\pm$0.1 & 2.6$\pm$0.2 & -4.4$\pm$0.3 & $-$              \\
34NW    &  WN7      & 12.8$\pm$0.1 & 3.0$\pm$0.2 & -4.6$\pm$0.3 & $-$              \\
16SE2   &  WN5/6    & 12.0$\pm$0.1 & 2.6$\pm$0.2 & -5.0$\pm$0.3 & GCIRS 16SE2      \\
7W      &  WC9      & 13.1$\pm$0.1 & 2.5$\pm$0.2 & -3.8$\pm$0.3 & GCIRS 7W         \\
7SE     &  WC9      & 13.0$\pm$0.1 & 2.6$\pm$0.2 & -4.0$\pm$0.3 & GCIRS 7SE        \\
13E4    &  WC9      & 11.7$\pm$0.1 & 2.8$\pm$0.3 & -5.5$\pm$0.4 & GCIRS 13E4       \\
\hline
\end{tabular}
\end{center}
\end{table*}


\section{Analysis of individual stars}
\label{results}

In this section, we give the results of the detailed analysis of
individual stars. We mainly focus on Wolf-Rayet and the so -called
``\ion{He}{i}'' stars since their spectra have large enough S/N ratios to
allow quantitative spectroscopy. OB supergiants are studied by means
of the average spectrum of 10 of them \citep{pgm06}. The global
methodology presented in Sect.\ \ref{method} is not repeated for each
star: we only give specific comments when necessary.

However, as a preamble, we would like to say a few words about the
behavior of \hei. It is well known that this line is extremely
sensitive to any detail of the modeling, and in particular to the
amount of UV radiation. Indeed, the upper level of \hei\ is directly
coupled to the ground state by a transition at 584 \AA.  Dramatic
improvement has been achieved in recent years in the prediction of
this UV radiation, mainly due to the inclusion of line-blanketing in
the models. However, as we will see in the following analysis, \hei\
is still poorly reproduced in several stars. Recently, \citet{paco06}
have shown that the radiation at 584 \AA\ was partly controlled by the
strength of two Fe~{\sc iv} lines: artificially changing the strength
of these lines improved the fit of optical singlet He{\sc i} lines. We
have tried the same kind of tests in the present study, but it turned
out that the resulting \hei\ line profiles were little changed. The
reason is partly that in the Wolf-Rayet stars studied here, \hei\ is
an emission line and is controlled by recombination processes, while
in the O stars analysed by \citet{paco06}, \hei\ is in absorption and
depends much more on pure radiative transfer effects. Besides, the
optical depth of the \ion{He}{i} 584 \AA\ line is large in Wolf-Rayet
stars, partly controlling the population of the \hei\ upper
level. This does not mean that \hei\ is not sensitive to the UV
continuum: as we will discuss in Sect.\ \ref{dis_z}, subtle blanketing
effects can significantly change the appearance of \hei. Having said
that, we now turn to the detailed study of individual GC stars.

\begin{sidewaystable*}
\begin{center}
\caption{Derived stellar and wind parameters. The typical errors are: $\pm$3000 K on temperatures ($\pm$6000 K for Ofpe/WN9 stars), $\pm$0.2 dex on \lL\ and $\log \dot{M}$, 100 \kms\ on terminal velocities and $\pm$30\%\ on abundances (except special cases; see comments on individual stars). \label{tab_res}}
\begin{tabular}{clrrrrrrrrrrrrrrrr}
\hline\hline
Star   & ST         & T$_{*}$ & \teff & \lL  & R$_{*}$ & R$_{2/3}$ & M$_{K}$ & log \mdot &f$_{\infty}$& \vinf & H/He & C/He  & X(N)   & M    & $\log Q_{H}$ & $\log Q_{\ion{He}{i}}$ \\
       &            & [K]     & [K]   &   & [R$_{\odot}$] & [R$_{\odot}$] &  & [\myr]    &  & [\kms]& \# & \# & & [M$_{\odot}$] &[s$^{-1}$]&[s$^{-1}$] \\
\hline 		   				        								     
34W    & Ofpe/WN9 & 23000   & 19500 & 5.5  & 35.9   & 49.3     & -6.11   & -4.88     & 1.0 & 650   & 4.0  & $-$   &  $-$   & $-$  & 48.32 & 47.64 \\
16NW   & Ofpe/WN9 & 20000   & 17500 & 5.9  & 59.1   & 75.8     & -6.80   & -4.95     & 1.0 & 600   & 5.0  & $-$   &  $-$   & $-$  & 48.04 & 47.35 \\
16C    & Ofpe/WN9 & 21500   & 19500 & 5.9  & 63.9   & 79.5     & -7.05   & -4.65     & 1.0 & 650   & 2.5  & $-$   &  $-$   & $-$  & 48.76 & 47.60 \\
33E    & Ofpe/WN9 & 20000   & 18000 & 5.75 & 63.9   & 75.9     & -6.85   & -4.80     & 1.0 & 450   & 4.0  & $-$   &  $-$   & $-$  & 48.26 & 47.60 \\
AF     & Ofpe/WN9 & 23000   & 21000 & 5.3  & 28.1   & 34.5     & -5.77   & -4.75     & 0.1 & 700   & 2.0  & $-$   &  $-$   & $-$  & 48.06 & 47.82 \\
15NE   & WN8      & 34500   & 33000 & 5.25 & 11.9   & 13.0     & -4.65   & -4.70     & 0.1 & 800   & 0.0  &$<$1 10$^{-4}$& 0.0137 & 12.6 & 48.56 & 47.47 \\
AFNW   & WN8      & 37000   & 33000 & 5.5  & 13.9   & 17.5     & -5.19   & -4.50     & 0.1 & 800   & 0.1  &$<$1 10$^{-4}$& 0.0326 & 18.6 & 49.14 & 47.60 \\
9W     & WN8      & 40500   & 32000 & 5.4  & 10.2   & 16.4     & -5.20   & -4.35     & 0.1 & 1100  & 0.1  &$<$5 10$^{-5}$& 0.0133 & 15.8 & 49.11 & 47.79 \\
7E2    & WN8      & 37500   & 34500 & 5.2  &  9.5   & 11.1     & -4.25   & -4.80     & 0.1 & 900   & 0.0  &$<$8 10$^{-5}$& 0.0137 & 11.7 & 48.88 & 47.70 \\
13E2   & WN8      & 29000   & 29000 & 6.1  & 44.7   & 45.0     & -6.55   & -4.35     & 0.1 & 750   & 0.1  &$<$3 10$^{-4}$& 0.0167 & 82.5 & 49.49 & 47.61 \\
7SW    & WN8/WC9  & 34500   & 33000 & 5.55 & 16.8   & 18.0     & -5.19   & -4.70     & 0.1 & 900   & 0.0  & 0.005 & 0.0135 & 18.6 & 49.15 & 47.66 \\
15SW   & WN8/WC9  & 39000   & 35000 & 5.1  &  7.9   &  9.7     & -4.36   & -4.80     & 0.1 & 900   & 0.0  & 0.013 & 0.0229 & 10.3 & 48.77 & 47.83 \\
AFNWNW & WN7      & 36500   & 28500 & 5.25 & 10.7   & 17.2     & -4.59   & -3.95     & 1.0 & 1800  & 0.1  &$<$1 10$^{-4}$& $-$ & 12.6 & 48.87 & 47.77 \\
34NW   & WN7      & 34000   & 33000 & 5.6  & 18.2   & 19.3     & -4.71   & -5.30     & 0.1 & 750   & 1.0  &1.5 10$^{-4}$ &  0.0069  & $-$ & 49.25  & 48.13 \\
16SE2  & WN5/6    & 53000   & 41000 & 5.45 &  6.4   & 10.5     & -5.03   & -4.15     & 0.1 & 2500  & 0.0  &$<$1 10$^{-4}$& $-$ & 17.2 & 49.24 & 48.51 \\
7W     & WC9      & 47500   & 39000 & 5.1  &  5.3   &  7.8     & -3.79   & -5.0      & 0.1 & 1000  & 0.0  & 0.06  &  $-$   &  10.3 & 48.86 & 47.80 \\
7SE    & WC9      & 44500   & 36500 & 5.15 &  6.4   &  9.5     & -4.03   & -4.90     & 0.1 & 1000  & 0.0  & 0.04  &  $-$   & 11.0 & 48.88 &  47.82 \\
13E4   & WC9      & 42500   & 37500 & 5.8  & 14.7   & 18.7     & -5.44   & -4.30     & 1.0 & 2200  & 0.0  & 0.02  &  $-$   & 45.0 & 49.56 & 48.47 \\
\hline
\end{tabular}
\end{center}
\end{sidewaystable*}

\subsection{Ofpe/WN9 stars}
\label{res_ofpe}

In this Section we present the analysis of five Ofpe/WN9 stars. We
deliberately exclude the stars IRS16SW and IRS16NE since they are
binaries \citep[or candidates, see][]{martins06}.

\subsubsection{IRS34W}
\label{34w}

IRS34W is an Ofpe/WN9 star and the faintest of the LBV candidates
identified in the Galactic Center \citep{paumard04}. \citet{trippe06}
also showed that it had recently experienced photometric variability
attributed to the formation of dust in material possibly ejected in a
LBV-like outburst.

\begin{figure}
\includegraphics[width=9cm]{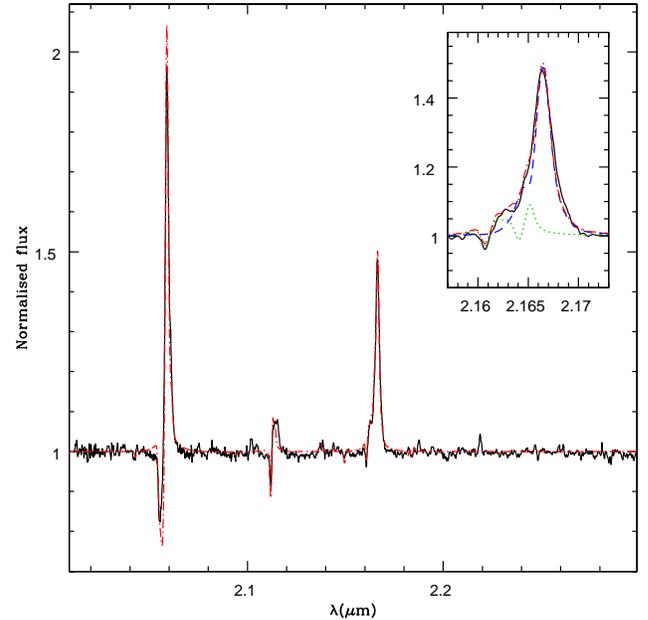}
\caption{Best fit (red dot-dashed line) of the observed K band spectrum of IRS34W (Ofpe/WN9, black solid line). The insert is a zoom on the \brg\ line, showing the contribution of H (blue dashed line) and \ion{He}{i} (green dotted line) to the total synthetic profile (red dot-dashed line). \label{fit_34w_K}}
\end{figure}

Due to the absence of a strong \teff\ indicator, we could find several
solutions to the fit of the K band spectrum by varying the He content,
the luminosity, and the mass loss rate for \teff\ in the range $\sim$
20000 - $\sim$ 33000 K. Fig.\ \ref{fit_34w_K} shows one of the
possible best fit models. For the acceptable effective temperatures,
the He/H ratio is larger than solar, in the range 0.25 - 0.6, while
the mass loss rate goes from $6 \times 10^{-6} - 2 \times 10^{-5}$
\myr\ and the luminosity is found in between $3$ and $6 \times\ 10^{5}
L_{\odot}$. It is important to note that these parameters are not
independent and only certain combinations within these ranges are
acceptable. For example, low \teff\ imply low luminosities, large
\mdot\ and large He/H ratio.

The morphology of \brg\ is well sampled in our observed spectrum.  In
particular, the blue side of the \brg\ emission line shows a
``shoulder'' and, at even shorter wavelengths, a small absorption
dip. This results from the combination of two effects: first, \brg\ is
a P-Cygni profile for which the blue absorption dip is partially
filled by emission; second, several He lines with both emission and
absorption profiles add to this \brg\ shape. The contribution of each
element is shown in the insert in Fig.\ \ref{fit_34w_K}. The exact
shape of this complex profile depends on the wind density in the line
formation region, and consequently depends on both $\beta$ and the
filling factor $f$ (in addition to the mass loss rate and the He/H
ratio). In order to fit this profile, we had to choose $\beta$ in the
range 2.0 -- 4.0, i.e. larger than 1, the value commonly found in O
supergiants. This is due to the fact that larger $\beta$ leads to
narrower profiles with stronger absorption dips and emission peaks
\citep[see][]{n81}. Similarly, the best fits were obtained with
\textit{unclumped} models. This is at odds with the current knowledge
that winds of massive stars are strongly inhomogeneous. However, since
this complex line profile depends on several parameters, we refrain
from concluding that the wind of IRS34W (and the other similar
Ofpe/WN9 stars) are homogeneous. More detailed investigations with
high resolution spectroscopy should help to better resolve this line
and improve our determination.

The only feature not well reproduced by our models is the emission at
2.112-2.115 \mum: it is wider in the observed spectrum than in our
models. The main reason is that this emission is a blend of several
lines, the identification of which is still under debate. The
concensus is that both \heid\ (blue part, through P-Cygni profiles)
and \niiia\ (red part) contribute to the emission. Clearly, the
\ion{He}{i} emission is present in our models but we are not able to
reproduce the \ion{N}{iii} emission. Actually, we are not convinced
that this emission is really due to \ion{N}{iii} since we do not
detect the lines \ion{N}{iii} 2.247 \mum\ and \ion{N}{iii} 2.251
\mum. \citet{geballe06} recently claimed that \ion{O}{iii} could be
responsible for that emission. Since the oscillator strength for this
transition is very uncertain, we decided not to fit the red part of
the emission complex at 2.112-2.115 \mum.

\subsubsection{IRS16NW}
\label{16nw}

\begin{figure}
\includegraphics[width=9cm]{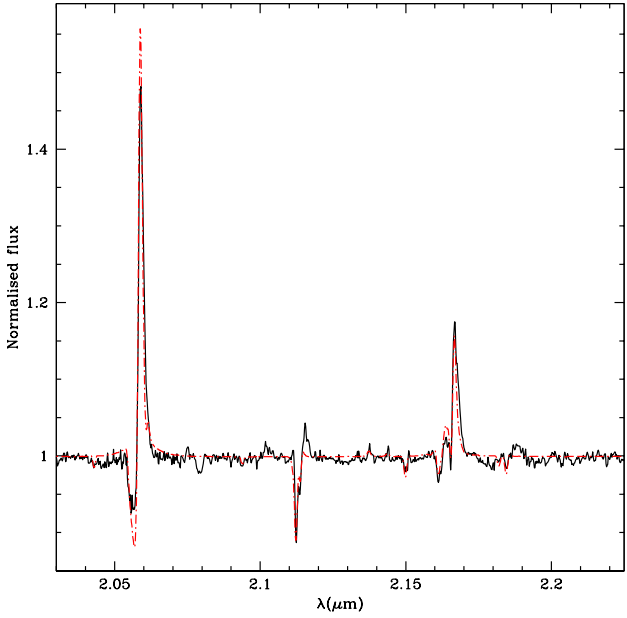}
\caption{Best fit (red dot-dashed line) of the observed K band spectrum of IRS16NW (Ofpe/WN9, black solid line).\label{fit_16nw}}
\end{figure}

The spectrum of IRS16NW is similar to IRS34W with the exception that
the emission part of the 2.112 \mum\ complex is weaker and the He
lines on the blue wing of \brg\ are mainly in absorption. However,
IRS16NW is 2.5 mag brighter, leading to a higher luminosity. The best
fit model of its K band spectrum is shown in Fig.\ \ref{fit_16nw}. For
this star too, the only problem is the emission at 2.112 \mum\ which
is absent in our model. As for IRS34W, large $\beta$ and unclumped
models gave the best fits.

Compared to the analysis of \citet{paco97}, we find a larger \teff,
although the range of values for which a fit can be achieved
encompasses their value. The luminosity is lower by a factor 2.4,
while \mdot\ is lower by a factor of 4.7. We also find a terminal
velocity smaller by 150 \kms. The main difference with \citet{paco97}
is the He content that we find larger than solar but smaller than the
H content. In our case, this indicates that IRS16NW has probably only
recently left the main sequence and is in an early evolved status.

\subsubsection{IRS16C}
\label{16c}

\begin{figure}
\includegraphics[width=9cm]{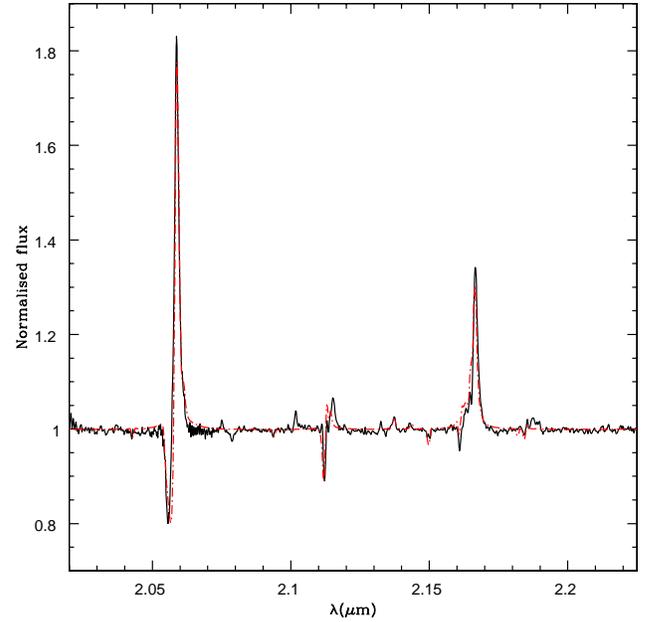}
\caption{Best fit (red dot-dashed line) of the observed K band spectrum of IRS16C (Ofpe/WN9, black solid line).\label{fit_16c_K}}
\end{figure}

IRS16C is the brightest star of our sample. It is one of the Ofpe/WN9
stars first discovered in the GC. As for the other stars of the same
spectral type, its \teff\ is poorly constrained. Fig.\ \ref{fit_16c_K}
shows one of the best fit models. A large value of $\beta$ as well as
an unclumped wind are favoured as for IRS34W. The feature at 2.115
\mum\ is not reproduced in our model, as well as the one at 2.100
\mum. These lines could be from \ion{N}{iii}, but the absence of
\ion{N}{iii} 2.247 \mum\ and \ion{N}{iii} 2.251 \mum\ emission weakens
this possibility. We detect a \ion{Mg}{ii} emission at 2.138
\mum. This emission is reproduced by our model for a twice solar Mg
content. However, this value should be interpreted with care given the
uncertainty in \teff\ and the weakness of the line.

As for IRS16NW, we find a lower He content and a lower mass loss rate
compared to the results of \citet{paco97}.

\subsubsection{IRS33E}
\label{33se}

\begin{figure}
\includegraphics[width=9cm]{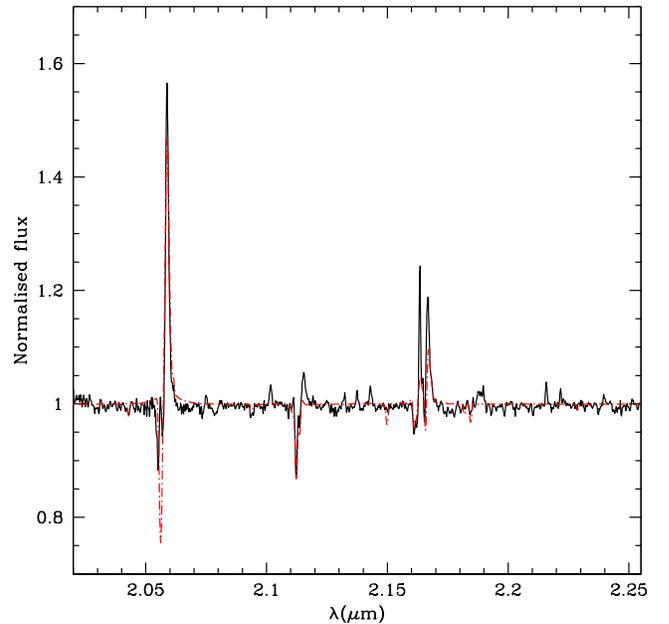}
\caption{Best fit (red dot-dashed line) of the observed K band spectrum of IRS33E (Ofpe/WN9, black solid line).\label{fit_33e_K}}
\end{figure}

The spectrum of IRS33E is similar to IRS34W, IRS16NW and IRS16C, and
so are the derived properties. The best fit model is shown in Fig.\
\ref{fit_33e_K}. As for IRS16C, the weak emission lines at 2.100 \mum\
and 2.115 \mum\ are not reproduced. Note that the structure of the
\brg\ emission is real and is observed at different epochs.

\subsubsection{AF star}
\label{af}

\begin{figure}
\includegraphics[width=9cm]{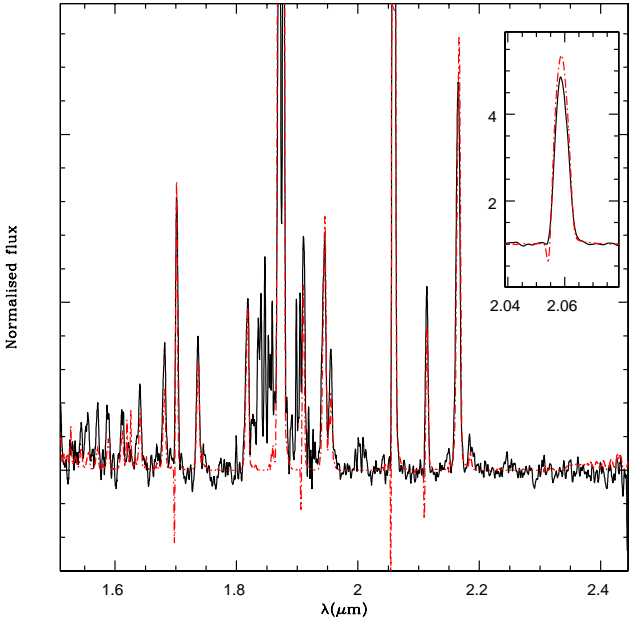}
\caption{Possible fit (red dot-dashed line) of the observed H+K band spectrum of the AF star (Ofpe/WN9, black solid line). Other combinations of \teff, He/H and \mdot\ lead to similar fits. The observed spectrum around 1.85 \mum\ is noisy and should be disregarded. See text for discussion. \label{fit_AF_HK}}
\end{figure}

The AF star is one of the first GC stars discovered and
analysed. Based on its SINFONI K band spectrum, \citet{forrest87},
\citet{ahh90} and \citet{paco94} classified it as a Ofpe/WN9 star.
However, compared to the other stars of the same type presented above,
AF has much broader lines. This indicates a stronger wind and possibly
a more advanced evolutionary state (see also Sect.\ \ref{dis_wn8}).
The best fit to the H+K band spectrum is shown in Fig.\
\ref{fit_AF_HK}. The effective temperature is poorly constrained due
to the absence of \ion{He}{ii} lines, as for the other Ofpe/WN9
stars. This implies a degeneracy between \teff\ and He/H: models with
$18000 K \lesssim \teff\ \lesssim 30000 K$ and $0.5 \lesssim \rm{He/H}
\lesssim 5.0$ gave reasonable fits (lower He/H being required at
larger \teff). Such behavior was already noted by
\citet{paco94}. We also found that \mdot\ $\sim 10^{-4}$ \myr\ were
necessary to fit emission at low \teff, while values of the order $1.5
\times 10^{-4}$ \myr\ were sufficient at high \teff. The luminosity is in
the range $1-2 \times 10^{5} L_{\odot}$. The values derived by
\citet{paco94} are consistent with our cool/He-rich/large \mdot\
models. The analysis of this star illustrates perfectly the degeneracy
one has to face when \teff\ is poorly constrained. In contrast to the
other Ofpe/WN9 stars analysed before, the shape of the \brg\ complex
is dominated by the wind, so that the contribution of the He lines
blueward of \brg\ is not resolved. Hence, the He content is poorly
constrained. In the comparison to the \citet{paco94} results, it is
also important to note that a different extinction was used. We
adopted $A_{\rm K} = 2.2$ while Najarro et al.\ had $A_{\rm K} =
3.0$. This influences the results, in particular the luminosity and
the mass loss rates (through their influence on M$_{\rm K}$, see
Sect.\ \ref{method}).

\subsection{WN stars}
\label{res_WN}

In the following, we present the results of the detailed study of 5
WN8, 3 WN7, 1 WN5/6 and 2 WN8/WC9 stars.For all the WN8 stars, the H
content is relatively low, but we cannot discriminate between H free
stars and stars with X(H) of a few percent. We thus quote a H mass
fraction $\lesssim$ 0.1.

\subsubsection{IRS15NE}
\label{15ne}

\begin{figure}
\includegraphics[width=9cm]{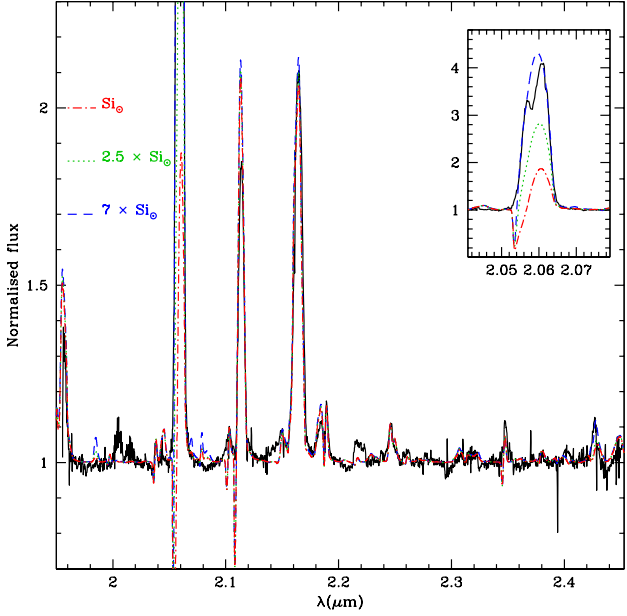}
\caption{Best fit (red dot-dashed line) of the observed K band spectrum of IRS15NE (WN8, black solid line). The different broken lines indicates models with similar parameters except the Si content. When it increases, the fits of Si~{\sc iv} 2.427 \mum\ and \hei\ are improved. See text for discussion. \label{fit_15ne}}
\end{figure}

Fig.\ \ref{fit_15ne} shows our best fit spectrum of the WN8 star
IRS15NE. The \ion{He}{ii} lines at 2.037, 2.189 and 2.346 \mum\ are
well reproduced and allow a good estimate of the effective
temperature. The fit of the \niii\ features is good, providing an
accurate N abundance determination. The main discrepancy concerns
the \hei\ line which is too weak. Reducing \teff\ improves the
fit but weakens the \ion{He}{ii} lines. The \ion{Si}{iv} feature at
2.427 \mum\ is also slightly too weak if a solar abundance is used for
Si. Fig.\ \ref{fit_15ne} shows the effect of increasing the Si content
by a factor of 2.5 and 7. Interestingly, the fit of the SiIV line
improves, as well as the fit of \hei! However, several weak Si lines
appear around 1.98 and 2.08 \mum. We do not detect these lines. Hence,
we cannot safely conclude that a super-solar Si abundance is required
for IRS15NE. The problem of Si~{\sc iv} 2.427 \mum\ in the initial
model may be due to incorrect atomic data for this line. In terms of
atmospheric structure, changing the Si content translates into a
slight variation of the temperature structure ($T$ is reduced in the
outer atmosphere) which is then responsible for the strengthening of
\hei\ (see also Sect.\ \ref{dis_z_hei}). This highlights once more
the extreme sensitivity of \hei\ to the very details of the modeling.

IRS15NE was previously studied by \citet{paco97} who found a lower
\teff, a larger luminosity and a larger mass loss rate. The terminal
velocity and He content were similar to the present value.

\subsubsection{AFNW}
\label{afnw}

\begin{figure}
\includegraphics[width=9cm]{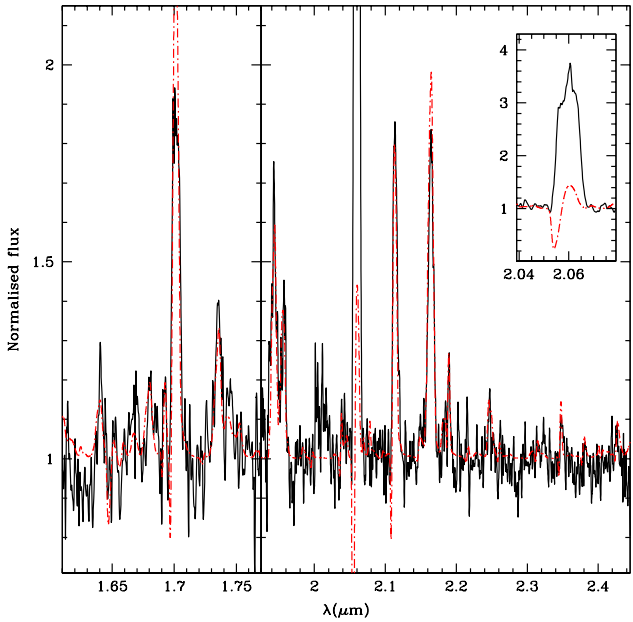}
\caption{Best fit (red dot-dashed line) of the observed H band (left) and K band (right) spectrum of AFNW (WN8). \label{fit_AFnw}}
\end{figure}

AFNW is located North West of the AF star and was assigned a spectral
type WN8 by \citet{pgm06}. The presence of both \ion{He}{i} and
He~{\sc ii} lines allows a rather robust \teff\ determination (see
Fig.\ \ref{fit_AFnw}). Only \hei\ is too weak and \ion{He}{i} 1.700
\mum\ too strong in our model.

\subsubsection{IRS9W}
\label{9w}

\begin{figure}
\includegraphics[width=9cm]{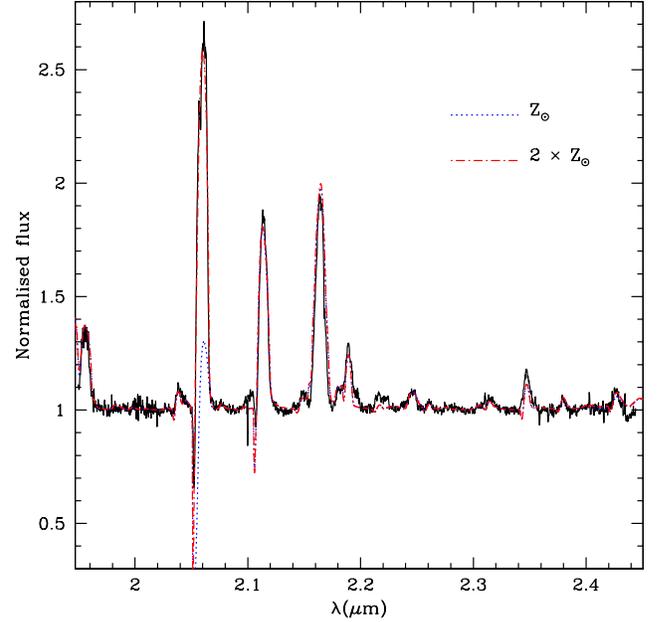}
\caption{Best fit of the observed K band spectrum of IRS9W (WN8, black solid line). The blue dotted line is a model with $Z = Z_{\odot}$ while the red dot-dashed line is for $Z = 2 \times\ Z_{\odot}$. Note the change of \hei\ with metallicity, tracing its extreme sensitivity to UV radiation. \label{fit_9w_K}}
\end{figure}

IRS9W is the WN8 star of our sample with the cleanest spectrum and the
strongest \ion{He}{ii} lines. Fig.\ \ref{fit_9w_K} shows that our best
model is able to perfectly reproduce most of the features, with the
notable exception of \hei. Given the quality of the observed spectrum
and the presence of several He{\sc i} and \ion{He}{ii} lines very well
reproduced by our model, we think the effective temperature is well
constrained (see in particular the ratio of \heii\ to
\heie). Decreasing \teff\ to strengthen \hei\ leads to weaker
\ion{He}{ii} 2.037, 2.189 and 2.346 \mum\ lines. Modifying the
luminosity and mass loss rate so that they still lead to a good match
of the observed K band magnitude and of the strength of emission lines
does not lead to any improvement as far as \hei\ is
concerned. Interestingly, the situation got better when we tried to
increase the global metallicity to a value of 2 times $Z_{\odot}$. In
that case, we could get a strong \hei\ line without degrading the fit
of the other diagnostics. This points once again to the extreme
sensitivity of \hei\ to UV opacities, a larger metallicity
corresponding to a softer UV radiation. A more detailed discussion of
the behavior of \hei\ as regards metallicity changes is given in
Sect.\ \ref{dis_z_hei}.

\subsubsection{IRS7E2}
\label{7e2}

\begin{figure}
\includegraphics[width=9cm]{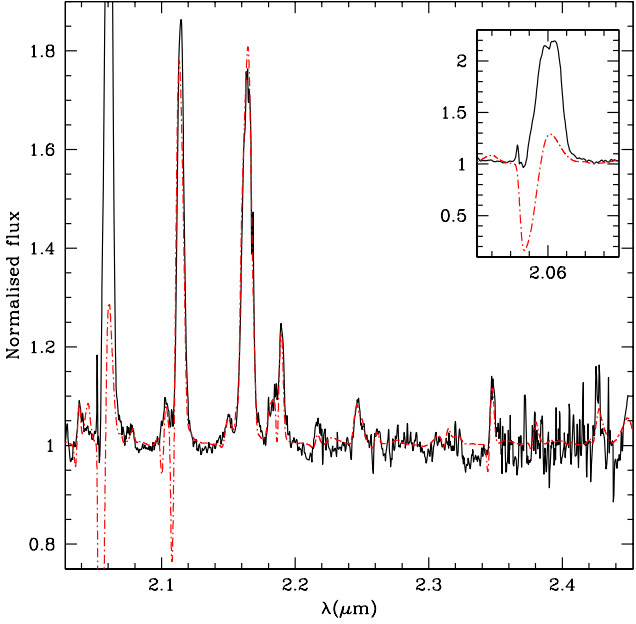}
\caption{Best fit (red dot-dashed line) of the observed K band spectrum of IRS7E2 (WN8, black solid line).\label{fit_7e2_K}}
\end{figure}

The best fit model to the K band spectrum of IRS7E2 is shown in Fig.\
\ref{fit_7e2_K}. All lines are well reproduced, except \hei. The
excellent fit of the \niii\ line allows a good N abundance
determination. \teff\ is also well constrained since \heii, \heiib\
and \heiic\ are clearly detected. As for IRS9W, we have tried to
increase the metallicity to improve the fit of \hei, but this time,
the line barely reacted and remained too weak.

\subsubsection{IRS13E2}
\label{13e2}

\begin{figure}
\includegraphics[width=9cm]{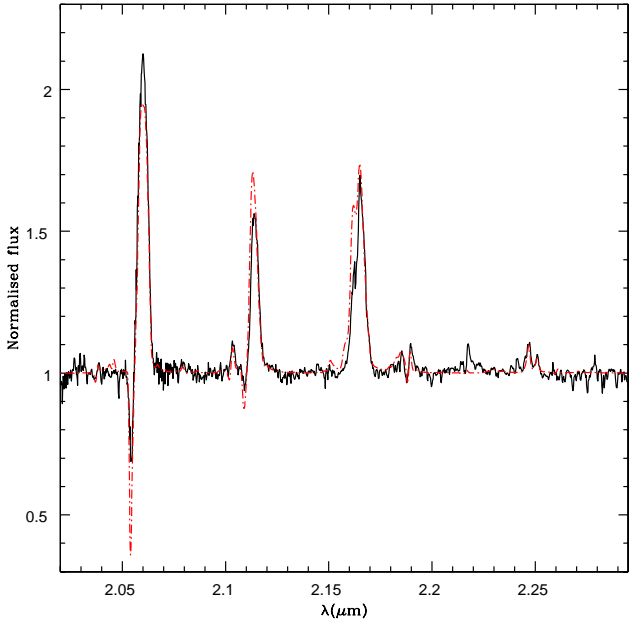}
\caption{Best fit (red dot-dashed line) to the observed K band spectrum of IRS13E2 (WN8, black solid line).\label{fit_13e2_K}}
\end{figure}

IRS13E2 is the brightest member of the IRS13E cluster and is
classified as WN8. Our best fit model is shown in Fig.\
\ref{fit_13e2_K}. An effective temperature of 29000 K was required to
fit the He spectrum. It is the most luminous WN8 star of our sample
However, the presence of dust in the IRS13E cluster may hamper our
determination. We will argue in Sect.\ \ref{res_WC} and
\ref{dis_irs13} that we probably derive only an upper limit on the
luminosity.  

\citet{paco97} already mentioned the stars of IRS13E in their study,
but at that time they could not resolve the components and analysed
the global spectrum. Hence, a direct comparison with their results is
meaningless.

\subsubsection{IRS7SW}
\label{7sw}

\begin{figure}
\includegraphics[width=9cm]{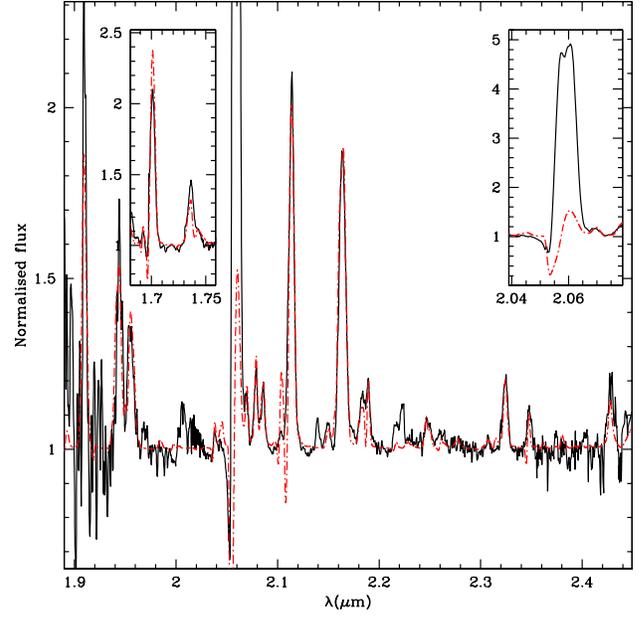}
\caption{Best fit to the observed H+K band spectrum of IRS7SW (WN8/WC9).\label{fit_7sw}}
\end{figure}

\citet{pgm06} classified IRS7SW as WN8, but in view of the present
results (see below), we refine its identification to WN8/WC9. Our best
fit model is presented in Fig.\ \ref{fit_7sw}. The preferred \teff\
allows a reasonable fit of the Carbon lines and of \heii, but seems a
little too low to account for \heiib\ and \heiic. However, increasing
\teff\ leads to a too strong \heii\ line. The presence of \ion{C}{iv}
and \ion{C}{iii} as well as \ion{N}{iii} lines indicates that IRS7SW
is likely a WN/WC star. This is confirmed by the abundance
determination (see also Sect.\ \ref{dis_wnc} for a quantitative
discussion). We note too unidentified lines: one feature at 2.140
\mum, and another one at 2.224 \mum. The former cannot be attributed
to \mgii\ since \teff\ is too high. Besides, \mgii\ is a doublet while
we clearly see only one component. The line at 2.224 \mum\ cannot be
due to \ion{Na}{i} since again \teff\ is too large. Interestingly,
these lines are also found in most of the WC9 stars of our sample, as
well as in the WN8/WC9 star IRS15SW (see next Section). We conclude
that they are typical of C-rich stars. Note that these lines are also
present and not identified in the WC9 stars of the \citet{figer97}
sample (see their Fig. 9).

\subsubsection{IRS15SW}
\label{15sw}

\begin{figure}
\includegraphics[width=9cm]{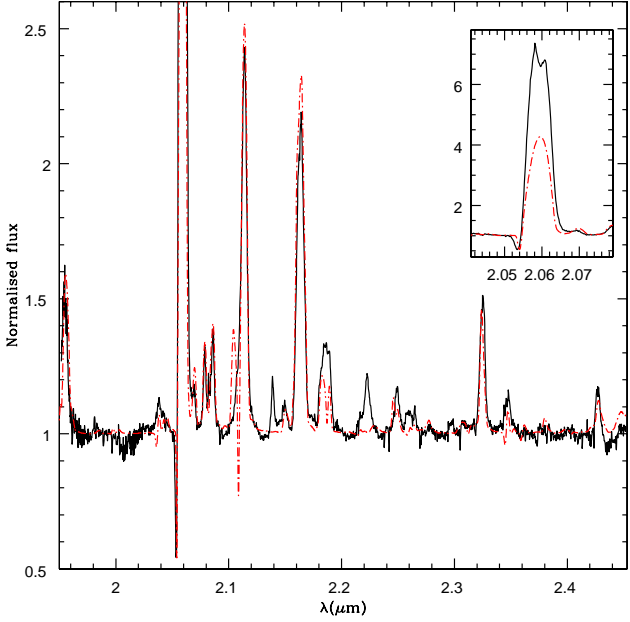}
\caption{Best fit (red dot-dashed line) to the observed K band spectrum of IRS15SW (WN8/WC9, black solid line).\label{fit_15sw}}
\end{figure}

IRS15SW is a late WN star showing C lines in its K band spectrum so
that it was classified as WN8/WC9 by \citet{pgm06}. Most of the lines
are reproduced by our best model (Fig.\ \ref{fit_15sw}). A notable
exception is the \ion{He}{i}/\ion{He}{ii} complex around 2.185
\mum. The emission is stronger than our model. Changing \teff\ does
not help since we fit either the blue or red side of the emission but
never the whole complex. Besides, \teff\ is relatively well
constrained by the other \ion{He}{ii} lines. We also do not perfectly
fit the blue absorption dip of \hei. This may require a larger \vinf,
but in that case the other emission lines are too broad. The C lines
are perfectly matched, allowing a reliable abundance (and \teff)
determination. IRS15SW will be further discussed in Sect.\
\ref{dis_wnc}.

IRS15SW was studied by \citet{paco97}. We find a much larger \teff,
essentially because we can rely on several \ion{He}{ii} lines and on
the \ion{C}{iv}/\ion{C}{iii} ratio (note the good fit of C lines in
Fig.\ \ref{fit_15sw}). We also find a much lower luminosity and a
lower clumping corrected mass loss rate. The terminal velocity and H
content are however similar (see Fig.\ \ref{fit_15sw} for a
complete view of the He and H spectrum).

\subsubsection{AFNWNW}
\label{afnwnw}

\begin{figure}
\includegraphics[width=9cm]{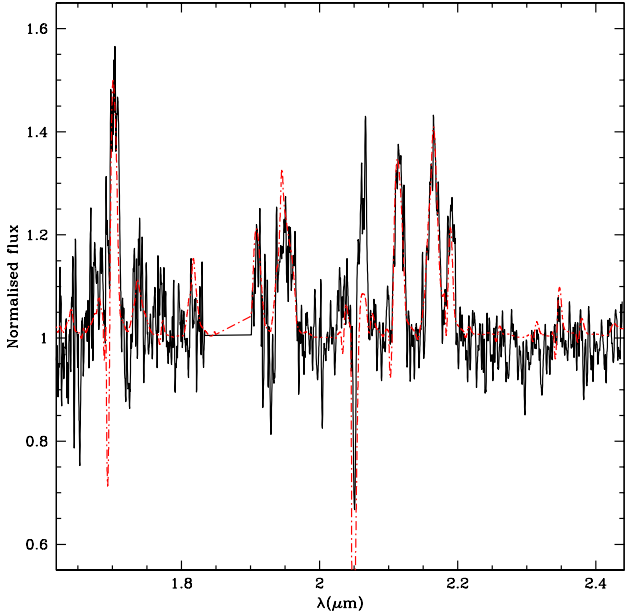}
\caption{Best fit (red dot-dashed line) to the observed H+K band
spectrum of AFNWNW (WN7, black solid line). The region around P$_{\alpha}$
was cut since it is not reliable. The observed spectrum was also
smoothed for clarity. \label{fit_afnwnw}}
\end{figure}

AFNWNW is a WN7 star. Most of its lines are reasonably well reproduced
by our best fit model (see Fig.\ \ref{fit_afnwnw}). The main problems
are the too weak \hei\ line and the too narrow \ion{He}{i}/\ion{He}{ii} complex at
2.18-2.19 \mum. Note however that the S/N ratio is rather low,
preventing an accurate determination of the physical parameters. The
absence of \ion{C}{iv} emission indicates an upper limit of 10$^{-4}$ for
C/He. Note that our best fit model is unclumped. We cannot derive the
clumping factor from the observed spectrum due to the poor S/N ratio.

\subsubsection{IRS34NW}
\label{34nw}

\begin{figure}
\includegraphics[width=9cm]{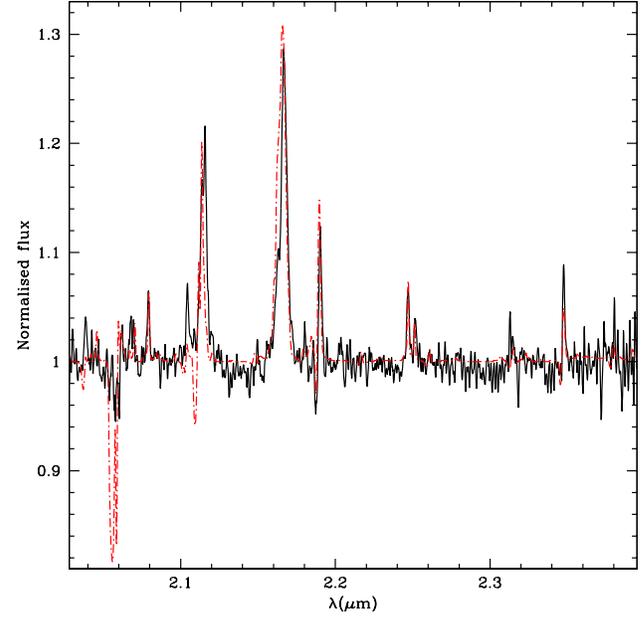}
\caption{Best fit (red dot-dashed line) to the observed K band
spectrum of IRS34NW (WN7, black solid line). \label{fit_34nw}}
\end{figure}

IRS34NW is a WN7 star. It has narrower and weaker lines than AFNWNW,
reflecting its weaker wind. The \ion{C}{iv} lines indicate a
slightly sub-solar C abundance (see best fit in Fig.\
\ref{fit_34nw}). Together with the N content derived from \ion{N}{iii}
lines, this reveals an early stage of CNO processing (compared to
AFNWNW which has no detectable \ion{C}{iv} lines). IRS34NW still has a
significant amount of Hydrogen. We conclude that IRS34NW is less
evolved than AFNWNW in spite of a similar spectral type. This likely
reflects different initial masses.

\subsubsection{IRS16SE2}
\label{16se2}

IRS16SE2 is the earliest WN star of our sample \citep[WN5/6,
see][]{pgm06}. Consequently, it also has the highest effective
temperature (41000 K) which is quite well constrained by the presence
of both \ion{He}{i} and \ion{He}{ii} lines in the K band
spectrum. Helium is indeed doubly ionised in most of the atmosphere,
except in the very outer part where it recombines, leading to the
\ion{He}{i} absorption trough near 2.044 \mum. This feature being due
to a blueshifted \hei\ absorption, we have a good estimate of the wind
terminal velocity (2500 \kms). The absence of \ion{C}{iv} emission
around 2.08 \mum\ -- expected for such a large \teff\ -- sets an upper
limit on the Carbon abundance ($C/He \lesssim 10^{-4}$ by number),
indicating CNO processing.

\citet{paul96} studied two WN6 stars with K band spectra very similar
to IRS16SE2. Their results are in excellent agreement with the present
ones: $T_{\star} \sim 55000 K$, $\lL \sim 5.4$ and $\log
\frac{\dot{M}}{\sqrt{f}} \sim -3.9$. Since \citet{paul96} found no
difference between their IR analysis and optical results, we are
confident that the parameters we derive for IRS16SE2 for pure IR
diagnostics are reliable.

\begin{figure}
\includegraphics[width=9cm]{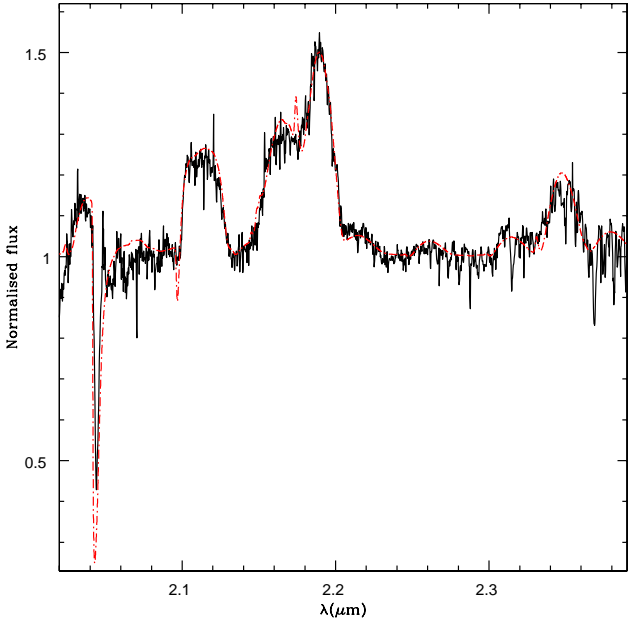}
\caption{Best fit (red dot-dashed line) of the observed K band spectrum of IRS16SE2 (WN5/6, black line).\label{fit_16se2_K}}
\end{figure}

\subsection{WC9 stars}
\label{res_WC}

Here, the stellar and wind parameters of three WC9 stars are
derived. One important word of caution is necessary though. WC9 stars
are often associated with dust \citep{williams87}. The origin of this
dust is not completely understood, but wind-wind interaction in binary
systems is the favoured mechanism. Observation of dust spirals (also
named 'pinwheels') around several WC9 stars strongly support this
scenario \citep{tuthill99}. Recent observations of the 'Cocoon stars'
after which the Quintuplet cluster was named showed such pinwheels
\citep{tuthill06}.
 
The presence of dust in WC9 stars complicates the analysis of their IR
spectra since it produces an additional emission which adds to the
stellar+wind continuum. In practice, if WC9 stars are analysed under
the assumption that they are dust-free, their continuum is
over-estimated. Consequently, when normalizing their spectra, lines
appear weaker. This implies under-estimates of the mass loss rates and
abundances. In addition, the derived luminosity is over-estimated
since the total continuum is composed of both the stellar+wind
continuum and the dust emission. Note however that \teff\ estimates
are less affected, since the ratio of lines is only weakly affected by
the presence of dust.

In the following, we discuss for each star the observational evidence
for dust and the reliability of the derived parameters.

\subsubsection{IRS7W}
\label{7w}

L-band observations of IRS7W were recently performed by
\citet{jihane05} \footnote{IRS7W is their WR2 star, and not WR1 as
they claim. Their WR1 star is IRS7SW.}. Inspection of their Table 1
shows that the colors of IRS7W are consistent with the extinction law
of the GC, at least in the HKL bands. A possible excess emission is
only seen in the M band due to a red L-M color. Consequently, we think
that our modelling, restricted to the H+K band spectrum, is not
hampered by any dust emission. The derived parameters can be trusted
within their error bars.

\begin{figure}
\includegraphics[width=9cm]{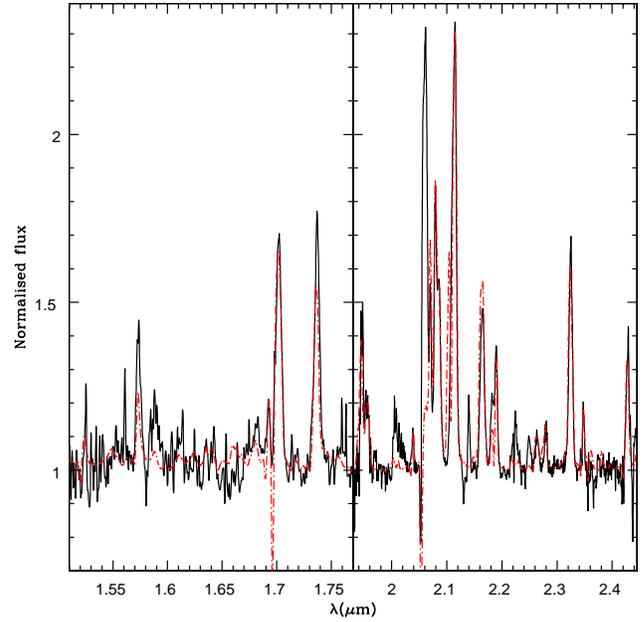}
\caption{Best fit (red dot-dashed line) of the observed H and K band spectra of IRS7W (WC9, black solid line).\label{fit_7w_HK}}
\end{figure}

Our best fit is shown in Fig.\ \ref{fit_7w_HK}. The effective
temperature is relatively well constrained by the presence of several
\ion{He}{ii} lines as well as \ion{C}{iv} and \ion{C}{iii}
features. The main discrepancy is once again the too weak \hei\ in
our model.  In addition, we note the two unidentified lines at 2.140
and 2.224 \mum\ as in WN/C stars.

\subsubsection{IRS7SE}
\label{7se}

\begin{figure}
\includegraphics[width=9cm]{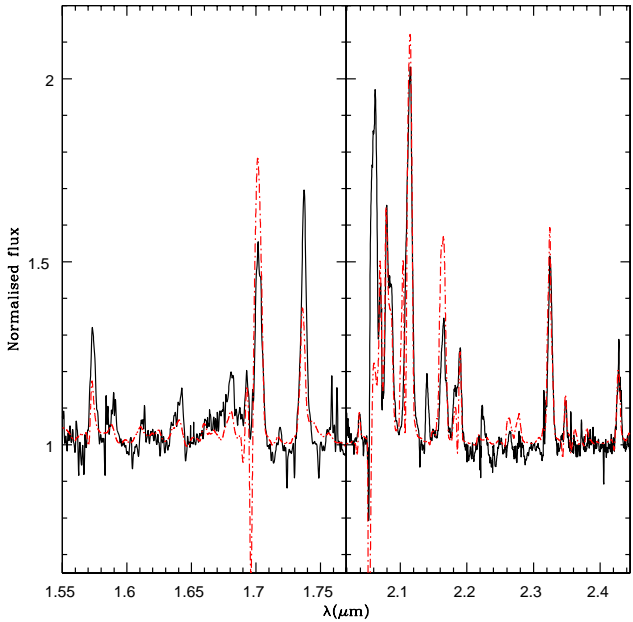}
\caption{Best fit (red dot-dashed line) of the observed H+K band spectrum of IRS7SE (WC9, black solid line).\label{fit_7se_HK}}
\end{figure}

IRS7SE is a WC9 star very similar to IRS7W. Unfortunately, we do not
have any information on its photometry (except for the K-band). We are
thus not able to check the possible contamination by dust. Adopting a
conservative approach, we consider that our results are only limits
(lower for \mdot\ and the C abundance, upper for \lL).  Fig.\
\ref{fit_7se_HK} shows a fit of similar quality compared to IRS7W,
with the same caveats (\hei, unidentified lines).

\subsubsection{IRS13E4}
\label{13e4}

\citet{maillard04} presented a detailed investigation of the stellar
content of IRS13E. Using HKL photometry, they report the discovery of
several very red sources in addition to the bright components IRS13E2
and IRS13E4. They interpret these sources as dusty Wolf-Rayet
stars. Unfortunately, IRS3E4 is not detected in their L band images,
so that the presence of a dust component in the spectrum of the WC9
star can not be tested. It is however interesting to note that
IRS13E2, the WN8 star analysed in Sect.\ \ref{13e2}, has a quite large
$K-L$ color. The entire cluster IRS13E is also known to be at the top
of a very prominent gas stream in L band \citep{clenet04}. Taken
together, these arguments indicate that the stars of IRS13E may well
all bathe in a continuum emission due to hot dust. We thus conclude
that their derived parameters must be regarded as only indicative (see
also Sect.\ \ref{dis_irs13}).

\begin{figure}
\includegraphics[width=9cm]{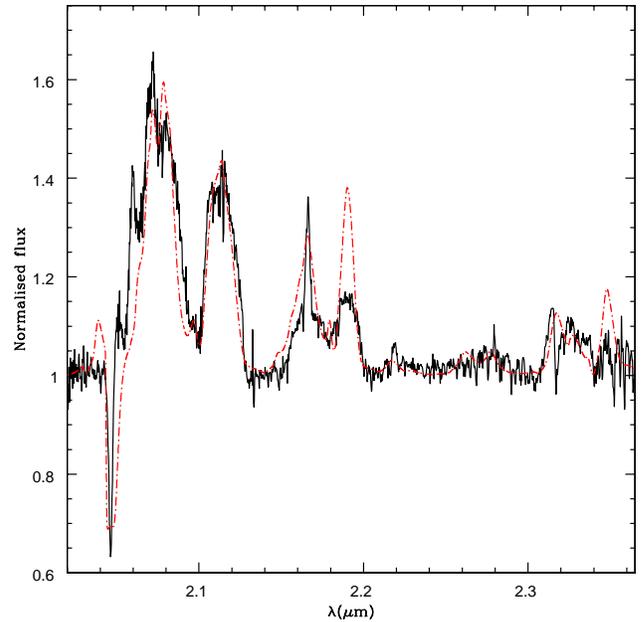}
\caption{Tentative fit (red dot-dashed line)of the K band spectra of the IRS13E4 (WC9, black solid line). See text for discussion of the discrepancies. \label{fit_13e4}}
\end{figure}

With this restriction in mind, we show a tentative fit of the spectrum
of the WC9 star IRS13E4 in Fig.\ \ref{fit_13e4}. The main C lines are
reasonably well reproduced by our model. However, the \heie\ and
\heii\ lines, while having the right line ratio, are too
strong. Reducing them would require a reduction of the mass loss
rate. This is at the cost of a reduction of the C lines (which could
be compensated by an increase in the C abundance), \textit{and} of a
modification of the shape of the lines. With lower \mdot, lines get
narrower and more centrally peaked because the density decreases. In
conclusion, and given the above discussion, we argue that the near-IR
spectrum of IRS13E4 is certainly contaminated by dust emission. Such a
component could explain that the \ion{He}{i} and \ion{He}{ii} lines
are too weak (being diluted). Carbon lines are also certainly diluted,
so that our C abundance estimate is likely a lower limit. We also
probably over-estimate the total luminosity.

\subsection{OB supergiants}
\label{OBIa}

Due to the rather limited signal to noise ratio of most of the spectra
of the OB stars known in the central cluster (S/N of the order of a
few tens on average), we have restricted ourselves to a general study
of the properties of these stars. For that, we have used the average
spectra of 10 supergiants presented in
\citet{pgm06}. Hence, we have derived average stellar and wind
parameters for this population of OB stars.

\begin{figure}
\includegraphics[width=9cm]{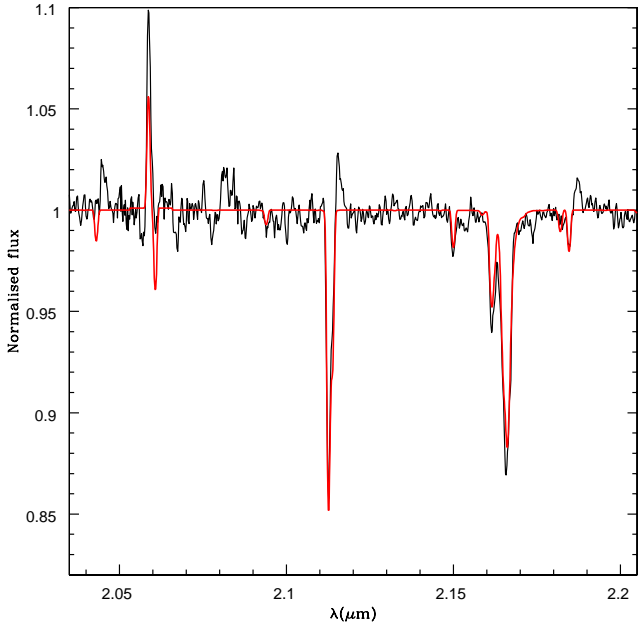}
\caption{Best fit (red solid line) of the average spectrum of the 10 supergiants \citep[see][]{pgm06}.\label{fit_OBIasum}}
\end{figure}

Fig.\ \ref{fit_OBIasum} shows the best fit model. The main parameters
used for this model are: \teff\ = 27500 K, \logg\ = 3.25, \lL\ = 5.33,
\mdot\ = $3 \times\ 10^{-7}$ \myr, \vinf\ = 1850 \kms, X(He) = 0.3 and
\vturb\ = 15 \kms. Additionally, a rotational velocity of $\sim$ 100
\kms\ could be estimated from the overall shape of all absorption
lines \footnote{For that, the model spectra were convolved to take
into account both the instrumental resolution and a rotational
broadening represented by a simple Gaussian function.}.  The effective
temperature is very difficult to constrain, just as in the Ofpe/WN9
stars studied previously. We can only put a constraint on the upper
limit of \teff\ from the absence of \ion{He}{ii} lines, especially
\heii. This upper limit is of around 32000 K. Giving a lower limit is
more challenging. We tried several values between 25000 and 32500
K. This range is thought to be appropriate since all stars
contributing to our average spectrum are late O / early B
supergiants. For these types of stars, \teff\ is expected to be around
25000 -- 30000 K \citep{msh05}. Reasonable fits could be obtained for
these different \teff. As we have no diagnostic to better constrain
\teff, we finally adopted 27500 K as a typical value. For the
luminosity, we also adopted \lL\ = 5.3 since this is typical of late O
supergiants \citep[e.g.][]{msh05}.

The mass loss rate is not strongly constrained either since most lines
are seen in absorption and do not appear to be filled by wind
emission. Here too, an upper limit on \mdot\ of a few $10^{-6}$ \myr\
can be given, above which \brg\ cannot be reproduced any more.

Finally, the He abundance is tentatively constrained from the
strength of \heid, \heif, \heig\ and \heie. The strength of these
lines not only depends on the He content, but also on the
microturbulent velocity. The slope of the velocity law (the $\beta$
parameter) plays a role too. As we have no independent way of
determining all these parameters, we tried different combinations with
reasonable values (10 $<$ \vturb\ $<$ 20 \kms, 1.0 $< \beta\ <$
2.0). In the end, we found that good fits could be achieved for He/H
ratios in the range 0.2-0.35. This value will be discussed in Sect.\
\ref{dis_ob}. Finally, we stress that we do not reproduce the 2.115
\mum\ emission. As discussed in Sect.\ \ref{34w}, this line is not
clearly identified and we did not try to fit it.

\section{Ionising radiation in the Galactic Center}
\label{dis_ion}

One of the crucial issues revealed by early studies of the Galactic
Center was the apparent incompatibility between the ionisation of the
gas and the ionising flux provided by the population of massive
stars. This was highlighted by \citet{paco97} and more recently by
\citet{lutz99} in a detailed analysis of ISO observations \citep[see
also][]{thornley00}. In view of our new quantitative analysis, we
argue that this issue is solved. Since this is an important result, we
give a detailed explanation of the different aspects of the problem
together with the proposed solutions.

\subsection{H and \ion{He}{i} ionising photons}
\label{Q_h_he}

The first part of the ``ionisation problem'' concerns the total number
of H and \ion{He}{i} ionising photons produced by the GC massive stars
(i.e. ionising photons short-ward of 912 \AA\ - \qh\ - and 504 \AA -
\qhei -). Nebular emission from the central H~{\sc ii} region
was used to constrain the various ionising fluxes. Radio measurements
of the free-free continuum by \citet{ekers83} showed that $\qh\ =
10^{50.4 \pm 0.3} s^{-1}$ \citep[see also][]{genzel94}. Later,
\citet{krabbe91} derived $\qhei\ \sim 10^{48.4} s^{-1}$ from a study
of the nebular \hei\ emission.  These measurements could then be
compared to the results of the quantitative modeling of the stellar
properties of the ``\ion{He}{i}'' stars by \citet{paco97}. Their conclusion
was that the ``\ion{He}{i}'' stars provided enough H ionising photons, but
failed by several orders of magnitude to produce the extreme UV flux
required to account for the nebular He ionisation. An underlying
population of stars hotter than the ``\ion{He}{i}'' stars but not detected was
suspected to be responsible for this harder flux.

Such a population has recently reported by \citet{pgm06}. Hence, a
re-estimate of the ionising power of the central cluster is
required. Table \ref{tab_res} gives \qh\ and \qhei\ for the stars
analysed in the present paper. In addition to these, \citet{pgm06}
identified 15 other evolved massive stars and 59 OB stars. To estimate
the total ionisation flux delivered by this population we used the
following approach: for the WC9 stars not analysed here, we adopted
the ionising fluxes of \citet{smith02} using their model WC number 3;
for the 2 Ofpe/WN9 stars IRS16NE and IRS16SW, we adopted the
parameters of IRS16C; for the WN7 stars not studied, we adopted the
values of AFNWNW, another WN7 star included in our sample; finally,
for the O stars, we adopted the calibration of \citet{msh05} (see
their Table 4). Since all supergiants have a spectral type between
O8.5 and B2 but some of them suffer from a classification uncertainty
of up to 2 spectral sub-types, we adopted the ionising flux of a O9.5I
star as typical of the GC supergiants. As for O dwarfs, the values of
\qh\ and \qhei\ of \citet{msh05} for a O9.5V star were chosen. For the
B dwarfs, we simply adopted a value ten times smaller than for OV
stars. This is a rough approximation which however has very small impact
on the final result, B stars providing a negligible ionising
flux. This leaves us with the following numbers: $\qh\ = 6.0 \times
10^{50} s^{-1}$ and $\qhei\ = 2.3 \times 10^{49} s^{-1}$ (or $\log \qh
= 50.8$ and $\log \qhei = 49.4$). The contribution of the different
classes of stars are gathered in Table \ref{tab_qi}. We thus conclude
that not only the H ionising flux but also the \ion{He}{i} ionising
flux required to reproduce the nebular emission can be provided by the
population of massive stars. In fact, they may even produce slightly
more ionising photons, indicating that the H~{\sc ii} region might be
density bounded. A final comment on the total \ion{He}{i} ionising
flux is needed. \citet{krabbe91} state that $\log \qhei = 48.4$
s$^{-1}$, but also that \qhei/\qh = 0.06. We find \qhei/\qh = 0.04, in
very good agreement. The difference in the absolute \qhei\ values
between our study and \citet{krabbe91} is their lower reference \qh\
($\log \qh = 49.6$).

At this point, a comment on the contribution of the ``\ion{He}{i}''
stars is necessary. \citet{paco97} showed that the 8 stars they
analysed could account for most of the ionising radiation of the
region. We see that with the current estimate, this conclusion would
not be valid. Why is that? The answer is rooted in 1) our lower
bolometric luminosities and 2) the inclusion of line-blanketing in the
atmosphere models. This ingredient was not available at the time of
the \citet{paco97} study. Test models reveal that in such stars,
line-blanketing effects lead to a large redistribution of the blocked
UV flux to longer wavelengths, mainly above 912 \AA. Consequently,
\qh\ is reduced significantly.

\begin{table*}
\begin{center}
\caption{Contribution of the different types of stars to the ionising fluxes. \label{tab_qi}}
\begin{tabular}{ccccccc}
\hline\hline
 Type of star & number of stars & \qh\                 & \%\ \qh\ total & & \qhei\              & \%\ \qhei\ total \\
              &                 & [$s^{-1}$]           &                & & [$s^{-1}$]          &                  \\
\hline                
OB V          & 23              & $3.1\times\ 10^{48}$ &    0.5         & & $2.6\times 10^{46}$ &   0.1 \\ 
OB I          & 30              & $3.0\times\ 10^{50}$ &   50.0         & & $6.7\times 10^{48}$ &  29.1 \\
Ofpe/WN9      & 8               & $2.5\times\ 10^{49}$ &    4.2         & & $3.4\times 10^{48}$ &  14.8 \\
WN            & 9               & $1.1\times\ 10^{50}$ &   18.3         & & $7.0\times 10^{48}$ &  30.4 \\
WN/C          & 2               & $2.0\times\ 10^{49}$ &    3.3         & & $1.1\times 10^{48}$ &   4.8 \\ 
WC            & 13              & $1.5\times\ 10^{50}$ &   25.0         & & $4.7\times 10^{48}$ &  20.4 \\
\hline
Total         &                 & $6.0\times\ 10^{50}$ &                & & $2.3\times 10^{49}$ &       \\
\hline
\end{tabular}
\end{center}
\end{table*}

\subsection{Ionising radiation and stellar population}
\label{Qi_evol}

The second important issue concerning the ionising radiation in the
Galactic Center was highlighted by \citet{lutz99}. In his study of
nebular fine structure mid-IR lines of metals observed with ISO, Lutz
pointed out that stellar evolution appears to fail to explain the GC
massive stellar population. This claim was based on computations of
population synthesis models for a single burst of star formation and a
standard Salpeter IMF. After 7 Myrs, the age of the population thought
to be appropriate at that time, the fraction of the total ionising
luminosity provided by the part of the HR diagram where the
``\ion{He}{i}'' stars are lying ($log T_{\rm star} < 4.5$ and $\L >
5.75$) was of the order of 1 \%. This was at odds with the results of
\citet{paco97} who argued that these stars could account for more than
half the total ionising luminosity. This discrepancy lead
\citet{lutz99} to the conclusion that stellar evolution - indirectly
tested here through synthesis population models - was not producing
enough cool stars or equivalently that the time spent by a massive
star in the cool part of its track was much too short.

This statement is no longer valid. Only six stars analysed here
have $log T_{\rm star} < 4.5$ and $\lL > 5.75$. And this is in the
case we include IRS13E2 for which we only have an upper limit on
\lL. To these four stars, we need to add the binary IRS16SW and the
binary candidate IRS16NE. Both stars likely have properties similar to
IRS16C. In total, the H ionising flux of these stars represents about 9
\%\ of the total \qh. This is an upper limit due to the possible
overestimate of IRS13E2's luminosity. If we exclude IRS13E2, the
remaining 7 stars contribute only 4 \% of $Q_{\rm H}^{\rm
total}$. This is in excellent agreement with what is expected from a
burst of star formation after 7 Myrs \citep[which is within the age
range now stated for the population, see][]{pgm06}. Once again, this
is mainly due to the recent discovery of a hot population of OB and
Wolf-Rayet stars responsible for the majority of the ionising
luminosity. Table \ref{tab_qi} shows that the OB supergiants and
Wolf-Rayet stars contribute more than 90 \% of the ionising flux. The
main conclusion is that standard stellar evolution -- used in
population synthesis models -- is able to account for the GC massive
stars.

\subsection{Mid-IR nebular Ne lines}
\label{mir_lines}

The third problem with the ionisation of the Galactic Center region
was also pointed out by \citet{lutz99} \citep[see
also][]{thornley00}. It concerned the low ionisation of the local gas
as derived from the ratio of fine structure mid-IR lines of different
ionisation states of Ne, namely \ion{Ne}{iii} 15.5 \mum\ and
\ion{Ne}{ii} 12.8 \mum. ISO observations revealed that
[\ion{Ne}{iii}]/[\ion{Ne}{ii}] was 0.05. Using the SED predicted by
the synthesis population model described in the previous section as an
input of a photoionisation model performed with the code CLOUDY
\citep{cloudy}, Lutz revealed that after 7 Myrs,
[\ion{Ne}{iii}]/[\ion{Ne}{ii}] was still as large as 1 to 2, or a
factor 50 to 100 more than the observed value.

To investigate this issue, we have performed photoionisation models
with CLOUDY (version C06.02). As an input SED, we have simply added
all individual SEDs computed for the stars presently studied. For
those stars which were not explicitly treated here, we have adopted
the SEDs of \citet{msh05} or those of similar stars present in the
current sample. Adopting the same density as \citet{lutz99} (3000
cm$^{-3}$) and our total ionising flux implies an ionisation parameter
$\log U = -0.6$ \footnote{the ionisation parameter is defined by $U =
\frac{Q_{\rm H}}{4 \pi r^{2} n c}$ where $r$ is the distance to the
ionising source (chosen to be 0.5 pc in our case) and $n$ is the
density.}. With these values, we obtain a ratio
[\ion{Ne}{iii}]/[\ion{Ne}{ii}] of 0.9-1.7 depending on the
geometry. This is still larger than the observed value (0.05). Note
however that if we use the ionisation parameter as \citet{lutz99}
($\log U = -1$), we get [\ion{Ne}{iii}]/[\ion{Ne}{ii}] $\sim$ 0.5 --
0.7. This is a factor 2 -- 3 lower than the values of Lutz.

How can we explain the still large values of
[\ion{Ne}{iii}]/[\ion{Ne}{ii}]? One possibility is that we
underestimate the density. There is evidence that values as large as
$10^{4-5}$ cm$^{-3}$ are required to produce [\ion{O}{i}],
[\ion{O}{iii}] and [\ion{Fe}{ii}] lines
\citep{genzel84,genzel85}. Using such large values (and the
corresponding ionisation parameter, $\log U = -1.2$ and $-1.6$),
CLOUDY models with our total stellar SED give
[\ion{Ne}{iii}]/[\ion{Ne}{ii}] $\sim$ 0.5 (for $10^{4}$ cm$^{-3}$) and
$\sim$ 0.1 (for $10^{5}$ cm$^{-3}$). This is in better agreement with
the observed ratio, although still a factor 2 -- 10 too
large. \citet{sf94} argued that the nebular spectrum of the Galactic
Center could be reproduced only if several gas components with
different densities were involved. In view of the present result, it
may well be that the Ne ionisation requires a large density material
(we can reproduce the observed [\ion{Ne}{iii}]/[\ion{Ne}{ii}] ratio
for a density of $3 \times\ 10^{5}$ cm$^{-3}$). 

Another explanation to the large theoretical
[\ion{Ne}{iii}]/[\ion{Ne}{ii}] ratio could be that we still overestimate
the flux at 41eV, i.e. the \ion{Ne}{ii} ionisation energy probed by
the [\ion{Ne}{iii}]/[\ion{Ne}{ii}] ratio. This part of the spectrum is
quite sensitive to blanketing effects. A slight increase in
metallicity could lead to a reduced flux and consequently a lower
[\ion{Ne}{iii}]/[\ion{Ne}{ii}] ratio \citep[e.g.][]{morisset04}.

One may also wonder which type of star contributes significant flux at
41eV. Actually, it turns out that the total SED is completely
dominated by a single star at this energy: the hot WN5/6 star
IRS16SE2. To test the influence of this star on the ionisation of the
GC gas, we removed its contribution to the total SED and ran test
CLOUDY models. Amazingly, for a density of 3000 cm$^{-3}$ (and $\log U
= -0.6$), the [\ion{Ne}{iii}]/[\ion{Ne}{ii}] ratio drops to 0.008,
less than the observed value!  This shows that this ratio is extremely
sensitive to the local radiation field. One can imagine that most of
the ionised gas is not illuminated by the IRS16SE2 radiation due to
shielding by local structures in molecular clouds. In that case, the
remaining radiation field is soft enough to maintain a low
[\ion{Ne}{iii}]/[\ion{Ne}{ii}] ratio.

In conclusion, one could say that a revised nebular modelling taking
into account both the spatial distribution of the gas and of the
ionising sources is required to solve the Ne ionisation problem. This
is well beyond the scope of the present paper.

\section{Stellar evolution in the Galactic Center}
\label{dis_evol}

In this Section we discuss, in view of the results of our quantitative
analysis, the evolution of massive stars beyond the main sequence. It
is generally accepted that stars in the mass range 25-60 \msun\ evolve
from O stars to WN H-poor stars through a LBV and/or red supergiant
phase before becoming WC stars (for $M > 40$ \msun). In the following,
we refine this scenario in the particular case of the GC, establishing
a plausible evolutionary sequence between Ofpe/WN9, WN8 and WN/C stars
(Sect.\ \ref{dis_ofpe}, \ref{dis_wn8}, \ref{dis_wnc}).  We also
quantitatively compare the position of the GC Wolf-Rayet stars in the
HR diagram to evolutionary tracks including rotation (Sect.\
\ref{dis_tracks}). Finally we discuss the properties of the IRS13E
cluster stars (\ref{dis_irs13}).

\subsection{Ofpe/WN9 stars}
\label{dis_ofpe}

In the present study, we have analysed five stars classified Ofpe/WN9
by \citet{pgm06}. Among these five stars, three have been studied by
\citet{paco97}: IRS16C, IRS16NW and AF. Compared to the Najarro et
al.\ analysis, we find a similar range of luminosities (although for
IRS16C our luminosity is lower) and the same terminal velocities
(within the uncertainty). The effective temperatures are higher in our
study, partly due to the inclusion of line-blanketing in our models as
already discussed. Nevertheless, \teff\ remains poorly constrained so
that the range of acceptable values overlap with the temperatures of
\citet{paco97}. As a consequence of the hotter \teff, our radii are
smaller. But the main differences concern 1) the mass loss rates and
2) the He content. Both parameters are linked to some extent: when
fitting \hei, \heid\ and \brg, adopting a larger He/H content will
require a larger \mdot\ in order to reproduce the level of
\brg\ emission. Of course, in that case \hei\ and \heid\ get stronger
too. But their absolute strength is also controlled by the He
ionisation which in turn depends on the effective temperature and the
line-blanketing effect. Since we used more realistic atmosphere models
as well as better spectra (higher S/N ratio and spectral resolution,
good correction from nebular emission), we argue that our derived
parameters represent an improvement over the result of
\citet{paco97}. In practice, we find values of \mdot\ 3 to 10 times
lower than Najarro et al., and much lower He contents (He/H $\sim$
0.2-0.5 compared to 1.3-3.0).

These revised parameters are important for assessing the evolutionary
status of the Ofpe/WN9 stars (see next Section). They are also very
interesting since they bring the GC Ofpe/WN9 stars closer to other
Galactic and LMC stars of this type. \citet{paul97} analysed a sample
of LMC Ofpe/WN9 stars and found that their He content was much smaller
than in the GC stars, which was tentatively attributed to
possible metallicity effects. Our new values are in better agreement
with the Crowther et al.\ measurement, showing that the LMC and GC
Ofpe/WN9 stars are chemically similar. In contrast, \citet{pasquali97}
found He/H $\sim$ 0.5 (by number) for a sample of LMC Ofpe/WN9 stars
partly overlapping with the Crowther et al.\ sample. These He contents
are only marginally larger than ours, and are certainly lower than the
values of \citet{paco97}. Concerning the mass loss rates, we find that
on average \mdot\ is systematically smaller than in the \citet{paul97}
and \citet{pasquali97} studies. This difference is surprising since
the lower metallicity of the LMC should lead to \textit{lower} mass
loss rates. Combined with the slightly larger He content, this may be an
indication of a more advanced evolution. Another indicator points to
the same conclusion: the terminal velocities of the GC Ofpe/WN9 stars
are usually larger than for stars of the same spectral type
\citep[see][]{pasquali97,bresolin02}. Interestingly enough, the only
Galactic star of this type studied by \citet{paul97} had a much larger
\vinf\ than the LMC stars. \citet{danielstsci} explained such a trend
by a more advanced evolution in the Ofpe/WN9 phase for stars in the
LMC, with the consequence of greater proximity to the Eddington limit
and implying a lower terminal velocity.

In conclusion, in view of our study, the class of Ofpe/WN9 seems to be
more homogeneous than previously believed. The difference between GC
and LMC Ofpe/WN9 stars is likely due to a different state of chemical
evolution: massive stars become Ofpe/WN9 stars later in a low Z
environment such as the LMC, mainly due to lower mass loss rates.

\subsection{An evolutionary link between Ofpe/WN9 and WN8 stars}
\label{dis_wn8}

On the basis of their K band morphology, all of the initial
``\ion{He}{i}'' stars have been classified as either Ofpe/WN9 or WN8
stars by \citet{pgm06}. These two spectral types are indeed
characteristic of relatively cool evolved massive stars. In the K
band, they are defined by strong \ion{He}{i} and \brg\ emission
lines. The relative intensity of these different lines is similar in
both spectral types: \hei\ is usually stronger than \brg, while \heid\
is weaker or of equal strength. The main differences are 1) the shape
of the lines (Ofpe/WN9 stars show P-Cygni profiles in the \ion{He}{i}
lines, while WN8 stars have strong pure emission lines), 2) their
width (WN8 stars have broader lines), 3) their absolute strength
(stronger lines in WN8 stars) and 4) the presence of weak He{\sc ii}
lines in WN8 stars. Inspection of Table \ref{tab_res} reveals that
quantitatively, these morphological differences are due to larger mass
loss rates (and to a lesser extent wind terminal velocities) as well
as higher effective temperatures for WN8 stars. Indeed, \teff\ ranges
from 30000 to 41000 K for WN8 stars, while they are lower than 30000 K
for Ofpe/WN9 stars. Mass loss rates are $\sim$ 2-4 times smaller in
Ofpe/WN9 stars. Luminosities are also slightly lower in WN8
stars. This comparison indicates that Ofpe/WN9 stars and WN8 stars may
be physically related and may well represent consecutive phases of a
single evolutionary sequence.

\begin{figure}
\includegraphics[width=9cm]{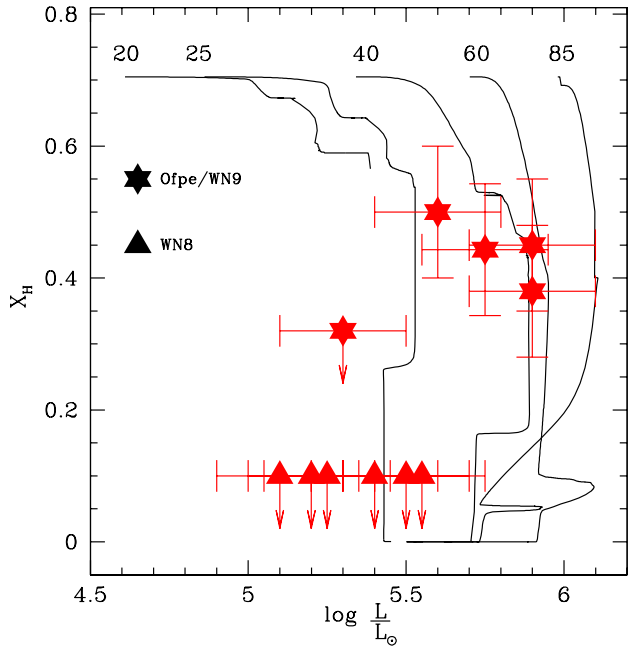}
\caption{Hydrogen mass fraction as a function of luminosity. Solid lines are the Geneva evolutionary tracks with rotation and $Z = Z_{\odot}$ \citep{mm05}. The ZAMS masses for each track are indicated. Stars show the position of the Ofpe/WN9 stars (IRS16C, IRS16NW, IRS33SE, IRS34W, AF) analysed here, while the triangles are the WN8 stars of our sample. The AF star lies at \lL\ = 5.3 and X(H)=0.3. See Sect.\ \ref{dis_tracks} for discussion. \label{XH_L}}
\end{figure}

Fig.\ \ref{XH_L} displays the H content as a function of luminosity in
evolutionary models (solid lines) and shows the position of the
Ofpe/WN9 and WN8 stars. It is clear that both types of stars gather in
different parts of the diagram: Ofpe/WN9 stars still show a
significant amount of hydrogen in their atmospheres whereas WN8 stars
are mainly H free. This can be interpreted as an evolutionary sequence
where Ofpe/WN9 stars evolve into WN8 stars. Such a scenario would be
consistent with the properties reported above. Ofpe/WN9 stars could
well be on the cool part of an evolutionary track. This track will
then loop back to the hot part of the HR diagram. As a star follows
this track, it evolves chemically, gets hotter and strengthen its wind
on its way to the WR phase. The hotter \teff\ and larger \mdot\
explains the appearance of He{\sc ii} lines and the stronger emission
lines, and the lower H content reveals the chemical evolution of the
star. In this scenario, the AF star could be in an intermediate
state. Its spectral morphology is similar to WN8 stars except that it
does not show He{\sc ii} lines, indicating a relatively cool \teff\
(confirmed by the quantitative analysis, see Table \ref{tab_res}). Its
mass loss rate is also more typical of WN8 stars. In Fig.\ \ref{XH_L},
the AF star lies in between the groups of Ofpe/WN9 and WN8 stars, with
a H mass fraction $\lesssim$ 0.3. Hence, it is also more evolved than
Ofpe/WN9 stars, but less than WN8 stars, and nicely fits in the
evolutionary scenario we suggest.

Another argument in favour of this scenario is the variability of both
types of objects. On the one hand, the Ofpe/WN9 stars in the Galactic
Center have been claimed to be LBV candidates, or even LBVs in a
quiescent phase \citep{paumard04,trippe06}. This is based on spectral
similarities between these stars and objects elsewhere in the Galaxy
and LMC known to be related to LBV stars. Besides, one of them --
IRS34W -- was shown to be photometrically variable on timescales of
months-years. \citet{trippe06} interpreted that as a sign of
obscuration by dust produced in material ejected in an LBV-type
outburst of the star. On the other hand, WN8 stars are known to be the
class of Wolf-Rayet stars experiencing the strongest variability
\citep{ant95,march98}. This high degree of variability may be related
to the LBV phenomenon. A link between WN8 stars and LBVs is also
favoured by the presence of LBV-like nebulosities around most of them
\citep{paul95_wnl}.

We thus argue that the GC Ofpe/WN9 stars are precursors of WN8 stars,
and are in a state closely related to the LBV phase. This picture is
fully consistent with the scenario of \citet{paul95_wnl}, further
extended by \citet{paul97}: a 25-60 \msun\ star evolves into a WN9-11
star \citep[similar to Ofpe/WN9, see][]{paul95_ofpe} before
experiencing a LBV phase and becoming a WN8 Wolf-Rayet star. The
properties of the WN8 stars analysed by \citet{paul95_wnl} and
\citet{herald01} are very similar to those of our sample WN8
stars. The only difference is a larger spread in H content: while we
find all WN8 to be almost H free, \citet{paul95_wnl} have both H free
stars and stars still showing a significant amount of
hydrogen. \citet{herald01} also found X(H) $\sim$ 20\%\ by
mass. However, this H content remains much lower than in the GC
Ofpe/WN9 stars. Hence, the suggested evolutionary scenario remains
valid.

\subsection{WN/C stars}
\label{dis_wnc}

WN/C stars are Wolf-Rayet stars showing both strong C and N lines
\citep{mg89,willis90}. Carbon is produced by the triple $\alpha$
reaction, while Nitrogen results mainly from CNO processing. It is
widely accepted that WN/C stars are core He burning stars with a
CNO-enriched envelope in which mixing processes have created a layer
with both H and He burning products. When mass loss reveals this
layer, the star turns into a WN/C star
\citep{langer91,mm03,mm05}. Quantitatively, such WN/C stars have C/N
ratios of the order 1.
 
For a long time, the fraction of WR stars in the WN/C state observed
in the Galaxy has been difficult to explain with evolutionary
models. Abundance profiles in such models were usually very steep in
the transition region between H and He burning products, so that the
region where both type of products were present was extremely
thin. Consequently, it was quickly removed by the stellar wind,
resulting in a very short lifetime of the WN/C phase. Consequently, the
number of WN/C stars predicted by such models was much lower than the
observed value. Significant improvements have been made in the last
years mainly due to the inclusion of mixing processes triggered by
rotation. \citet{mm03} have shown that rotation created shallower
abundance gradients in stellar interiors, increasing the size of the
mixed H and He burning products. As a consequence, the lifetime of the
WN/C phase is lengthened, resulting in a total number of WN/C stars in
better agreement with observations.

\begin{figure}
\includegraphics[width=9cm]{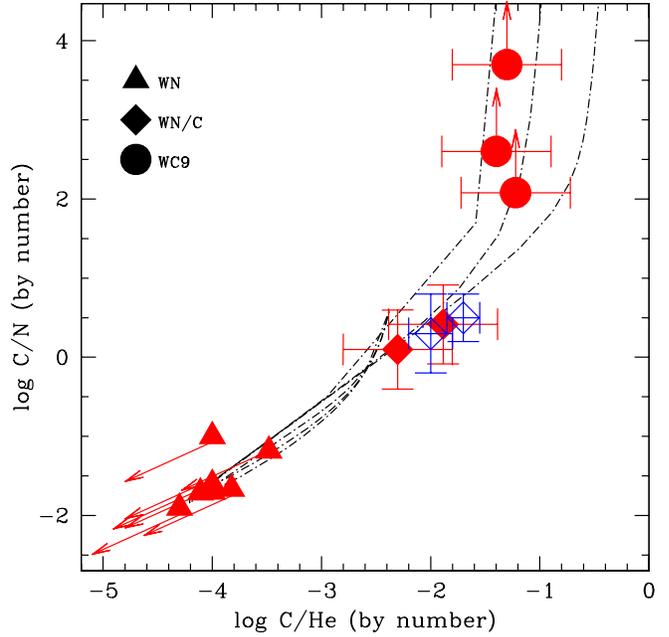}
\caption{$\log \rm{C/N}$ as a function of $\log \rm{C/He}$ (by number) from evolutionary tracks with rotation (dot-dashed line) from \citet{mm05}. Filled symbols are the WR stars analysed in the present paper (triangles: WN stars; diamonds: WN/C stars; circles: WC9 stars). Open diamonds are the WN/C stars WR8 and WR145 from \citet{paul95_wnc}. IRS7SW and IRS15SW, the two WN/C stars of our sample, have C/N and C/He ratios very similar to WR8 and WR145. The GC WN stars all lie at $\log \rm{C/He} \lesssim\ -3$. For these stars, we have only upper limits on the C abundance, and consequently on $\log \rm{C/N}$ and $\log \rm{C/He}$. The WC9 stars analysed here have large C/He ratios and only lower limits on C/N since no N lines are present in the spectra and only an upper limit on the N abundances can be estimated. \label{CN_CHe}}
\end{figure}

What about the GC Wolf-Rayet population? Our analysis of IRS15SW and
IRS7SW has revealed that they were significantly enriched in Carbon
compared to other WN stars, leading to a classification as WN8/WC9 stars. Fig.\ \ref{CN_CHe} shows
the position of our sample stars in a $\log \rm{C/N} - \log \rm{C/He}$
diagram. It turns out that IRS15SW and IRS7SW lie in between WN stars
(stars with low C content) and the C-rich WC9 stars. Their position is
also in excellent agreement with WR8 and WR145, two WN/C stars studied
by \citet{paul95_wnc}, and with the prediction of evolutionary
models. Hence we have a quantitative confirmation that IRS15SW and
IRS7SW are core He burning objects on their way to a WC
phase. \citet{mm03} argue that only stars with initial masses in the
range 30-60 \msun\ go through the WN/C phase: more massive stars have very
strong winds that quickly remove the CNO enriched envelope;
lower mass stars have a too H-rich envelope and He burning products
are too diluted. Estimates of the present-day masses of IRS15SW and
IRS7SW from the mass luminosity relation of \citet{hl96} for H free WR
stars gives 10.3 and 20.0 \msun\ respectively, implying that these
stars have lost 30 to 80 $\%$ of their mass through stellar winds.

If we compare the properties of IRS15SW and IRS7SW to WR8 and WR145
\citep[see][]{paul95_wnc}, the luminosities are all similar, in the
range $10^{5.1-5.5}$ \lsun, which in turn implies that the present-day
mass of these four objects are also very close (10 - 20 \msun). The
clumping corrected mass loss rates are also similar, with $\log
\dot{M}/\sqrt{f} \sim\ -4.3$. However, the terminal velocities of the
GC stars are smaller than for WR8 and WR145 (700-800 \kms\ as opposed
to 1390-1590 \kms) as well as the stellar temperatures (32-37 kK vs
41-48 kK). These differences reflect the different spectral types of
the two samples: while the GC WN/C stars are late type WR stars
(WN8/WC9), WR8 and WR145 are earlier \citep[WN6/WC4,
see][]{paul95_wnc}. Earlier WN and WC type stars have higher effective
temperatures and larger terminal velocities (see Fig.\
\ref{vinf_wr}). This can explain the observed trend: both types of
stars (early and late WN/C stars) have the same luminosity but early
WN/C stars are hotter, which implies that their radius is
smaller. Consequently, their escape velocity, scaling as
$(M/R)^{0.5}$, is larger (the present mass being approximately the
same). Since the terminal velocity is directly proportional to the
escape velocity, one naturally finds larger \vinf\ in early WN/C
stars.

We argue that the GC WN/C stars are most likely the descendents of WN8
stars. Their spectral morphology is extremely similar, except for the
presence of \ion{C}{iv} and \ion{C}{iii} lines in the WN/C stars. The
quantitative analysis confirms their ``twin'' character: the ranges of
values for \teff\, luminosities, mass loss rates, terminal velocities,
He and N content are the same for both WN8 and WN/C stars (see Table
\ref{tab_res}). The only possible exceptions are IRS13E2 (but see
discussion in Sect.\ \ref{dis_irs13}) and IRS9W. Fig.\ \ref{CN_CHe}
shows that the WN/C stars are more chemically evolved than WN
stars. Hence, we conclude that most of the GC WN8 stars will go
through a WN/C phase as two of them are currently doing.\\

Summarizing the last three Sections, we argue that in the Galactic
Center, stars with initial masses in the range 30 -- 60 \msun\ follow
the evolutionary sequence

\begin{center}
(Ofpe/WN9 $\rightleftharpoons$ LBV) $\rightarrow$ WN8 $\rightarrow$ WN/C
\end{center}

\noindent on their way to the supernova explosion. On the main
sequence, they probably appear as mid/early O stars. After the WN/C
phase, they most likely become late WC stars. Indeed the effective
temperatures of WN/C stars are slightly lower than the WC9 stars of
our sample. Hence, we can expect them to enter the WC sequence from
the low ionisation (or equivalently the low \teff) side and appear as
WC9 stars. The (Ofpe/WN9 $\rightleftharpoons$ LBV) sequence in the
suggested scenario indicates that Ofpe/WN9 and LBV stars are closely
inter-related and most likely represent different states of a single
evolutionary phase \citep[see discussion in][]{trippe06}.

\subsection{Are stellar evolutionary tracks too luminous?}
\label{dis_tracks}

The direct comparison between derived parameters and evolutionary
tracks by means of a classical HR diagram ($\log L - log \teff$) is
not straightforward for Wolf-Rayet stars.  In evolutionary models, the
``effective'' temperature (noted T$_{\rm evol}$ in the present
discussion) is defined at the outer boundary by Eq.\
\ref{eq:Tstar}. However, there is no atmosphere in such models and
strictly speaking, the radius in this case (noted R$_{\rm evol}$) is
not the one for which the optical depth is equal to 2/3 (exact
definition of \teff). It corresponds more to a hydrostatic radius. In
Wolf-Rayet stars, such a radius -- for which the atmosphere is
quasi-static -- is at an optical depth much larger than 2/3. Said
differently, R$_{2/3}$ for which \teff\ can be defined is much larger
than R$_{\rm evol}$. Consequently, \teff\ is lower than T$_{\rm
evol}$. To overcome this problem, one usually defines a radius
R$_{\star}$ (and the corresponding temperature T$_{\star}$, see Eq.\
\ref{eq:Tstar}) at an optical depth equal to 20. Such a radius is more
comparable to R$_{\rm evol}$ so that T$_{\star}$ and T$_{\rm evol}$
can be directly compared.

\begin{figure*}
 \centering
 \begin{minipage}[b]{8.8cm}
   \includegraphics[width=9cm]{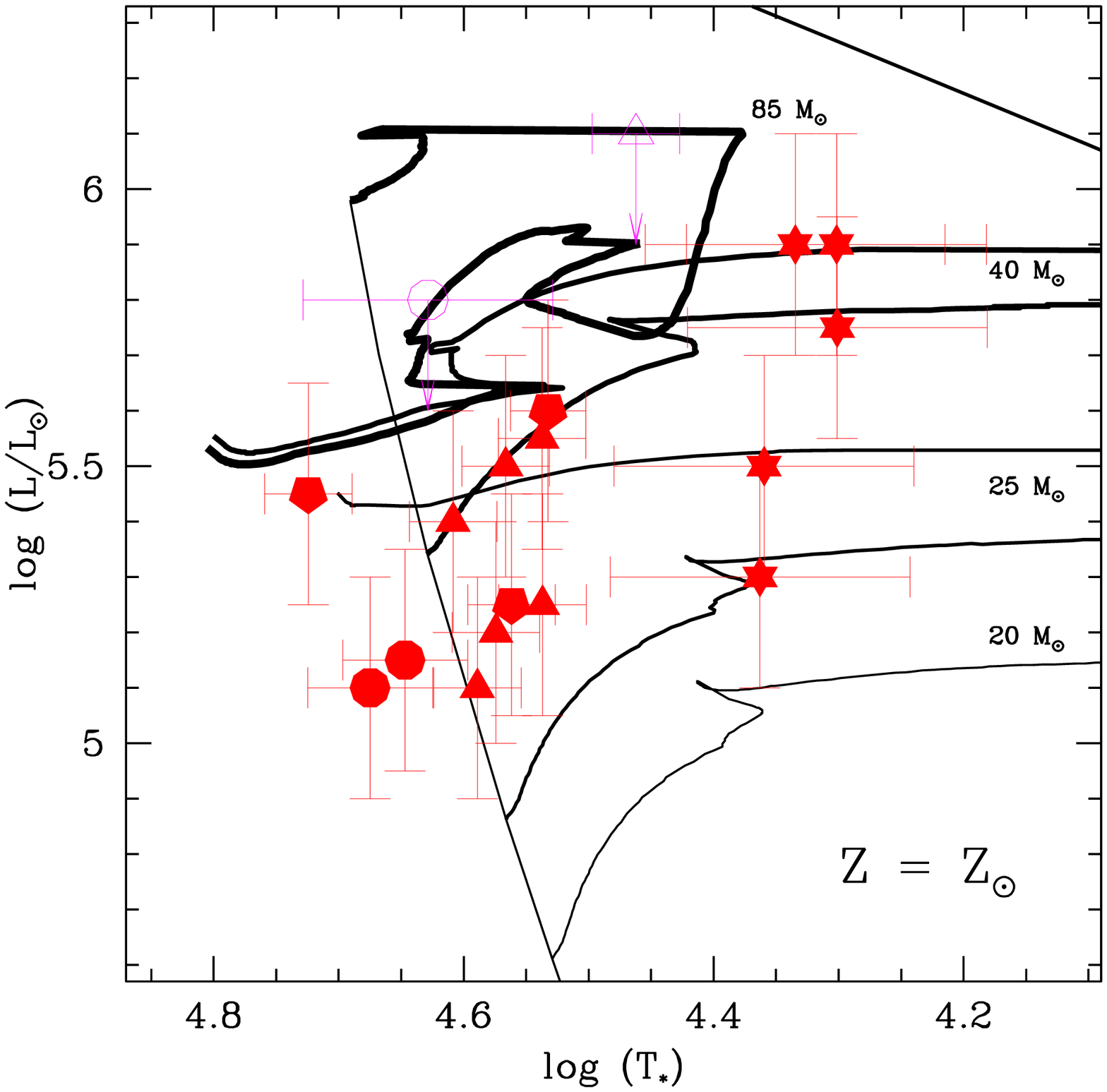}
 \end{minipage}
 \begin{minipage}[b]{8.8cm}
   \includegraphics[width=9cm]{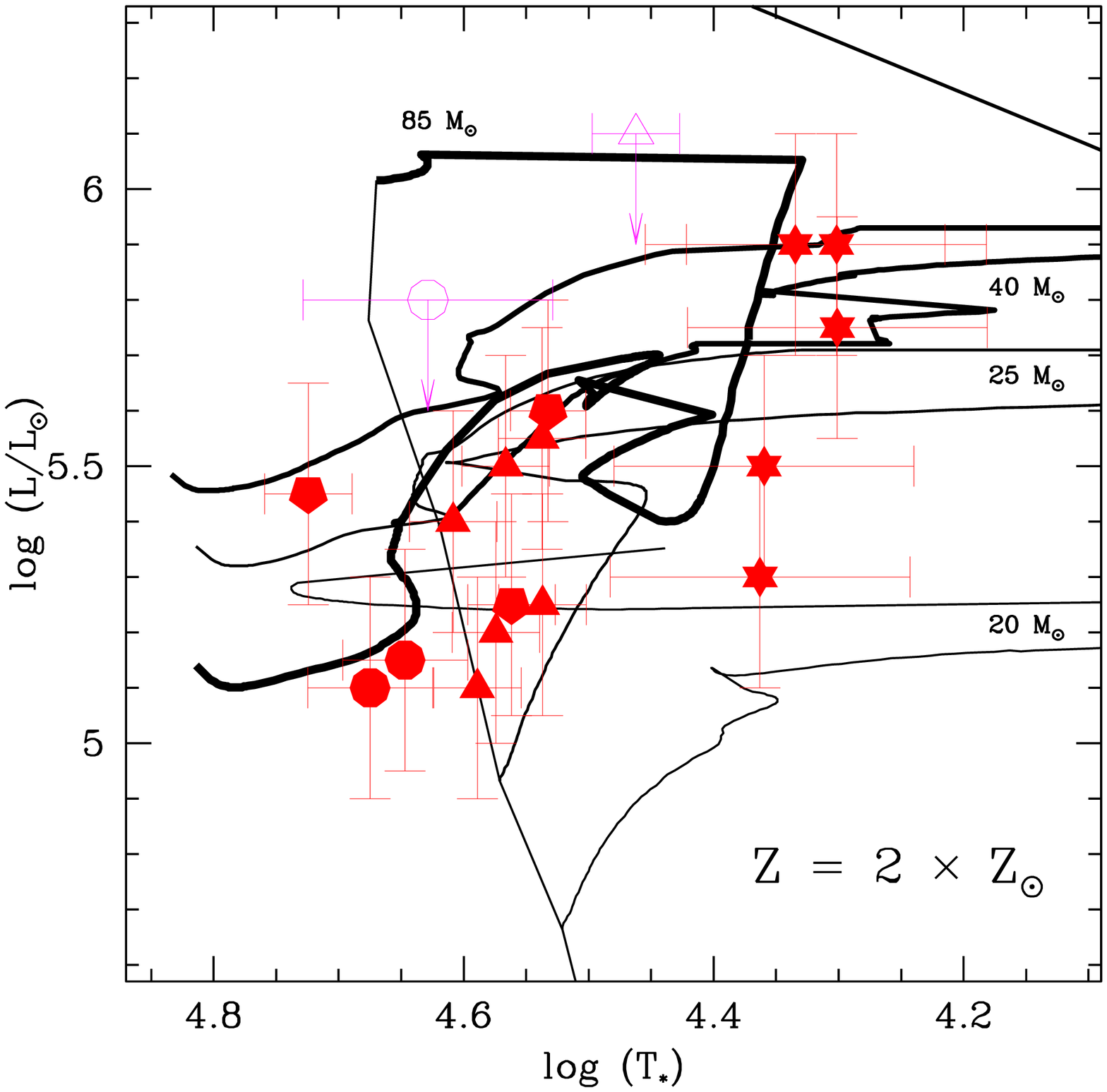}
 \end{minipage}
\caption{HR diagram of WR stars in the GC. \textit{Left}: solar metallicity tracks with rotation (\vsini\ = 300 \kms\ on the main sequence). \textit{Right}: twice solar metallicity tracks including rotation. Star symbols are Ofpe/WN9 stars, triangles WN8 stars, pentagons WN5-7 stars and circles WC9 stars. Different line thickness indicates different evolutionary tracks (the thicker the line, the higher the mass). ZAMS masses are marked for each track. The Humphreys-Davidson limit is also shown in the right upper part of each diagram.}
\label{hr_diag}
\end{figure*}

In Fig.\ \ref{hr_diag}, we show the HR diagram built using T$_{\star}$
for our program stars (See Table \ref{tab_res}). They are shown by the
filled symbols (see next Section for a discussion of the two stars
represented by open symbols). The evolutionary tracks including
rotation of \citet{mm05} for solar metallicity (left) and twice solar
metallicity (right) are overplotted. In the diagram at $Z =
Z_{\odot}$, we clearly see that the Ofpe/WN9 stars populate the
coolest region, while the more evolved Wolf-Rayet stars are hotter and,
on average, less luminous. Qualitatively, this trend is similar to the
40 \msun\ track: coming back from its redward extension, this track
goes to lower temperatures. However, this track, as well as the ones
for other masses, always remains at $\log L/L_{\odot} > 5.4$, in
contrast to most of the Wolf-Rayet stars of our sample which have $
5.1 < \log L/L_{\odot} < 5.5$. Hence, we see that quantitatively the
solar metallicity evolutionary tracks do not seem to explain the
evolution of the GC Wolf-Rayet stars. Note that this is independent of
any definition of the temperature, since the problem is due to
luminosities.

If we now focus on the HR diagram where the $Z = 2 \times\ Z_{\odot}$
metallicity tracks are plotted, we see that lower luminosities are
reached. This is due to the stronger mass loss rates in these tracks
which ``peel off'' the star more quickly, reducing its radius and
consequently its luminosity. Does that mean that the GC stars have a
supersolar metallicity? We will see in Sect.\ \ref{dis_z} that the
question is still largely debated. The position of the GC Wolf-Rayet
stars in the HR diagram may be an indication of a super-solar
environment. But this can also be due to inadequate mass loss rates in
evolutionary calculation at solar metallicity. If the amount of mass
lost during the Wolf-Rayet phase is underestimated in such
calculations, the tracks will be too luminous. Large \mdot\ can be
produced by a fast rotation \citep[see][]{mm00}, but the analysis of
OB stars average spectrum do not point to any particularly high
\vsini\ (see Sect.\ \ref{dis_ob}). As long as the GC metallicity
remains poorly constrained, a quantitative test of evolutionary tracks
and their input parameters, such as mass loss rates, is thus not
feasible. We will see in Sect.\ \ref{dis_z} that more work is needed
to narrow the range of acceptable values for Z in the Galactic Center.

\subsection{The stars of IRS13E}
\label{dis_irs13}

In the previous Section, we have deliberately excluded two stars from the
discussion: the ones represented by open symbols in Fig.\
\ref{hr_diag}. These stars are the two Wolf-Rayet components of the
cluster IRS13E, namely IRS13E2 and IRS13E4. The
reason is that they show peculiar properties compared to the bulk of
the GC Wolf-Rayet stars. Take first IRS13E2. It is a WN8 star (see its
spectrum in Fig.\ \ref{fit_13e2_K}) with parameters in marginal
agreement with the other WN8 stars of our sample, with the notable
exception of its luminosity ($\log L/L_{\odot} = 6.1$) which is
much larger than the range of values for WN8 stars \citep[5.1-5.5, see
Table \ref{tab_res} and][]{paul95_wnl}. The same statement is true
for IRS13E4. It is a WC9 star but exceptionally bright: while most
stars of this type have $\log L/L_{\odot} = 5.0$ \citep[see an example
in][]{paulwc9}, IRS13E4 is nearly 10 times more luminous. In addition,
it shows a large, although not unprecedented, terminal velocity (see
Fig.\ \ref{vinf_wr}).

How can we interpret these results? First, the most obvious
explanation is binarity. IRS13E was initially thought to be a single
star \citep{krabbe91} before being resolved into several components
\citep{maillard04}. The cluster is extremely compact and it is quite
possible that the bright components might actually be multiple
stars. But even with better spatially reduced images, it may not be
sufficient to distinguish between the components of a massive
binary. Spectrophotometric monitoring should then be used to test the
binary hypothesis. So far, IRS13E2 and IRS13E4 (Fig.\ \ref{fit_13e2_K}
and \ref{fit_13e4}) do not show any sign of double line spectrum. In
addition, photometric observations over the recent years (but with a
very coarse sampling) show a flat light-curve for both stars (Trippe
et al., in prep.). Better monitoring is required, but so far no clear
evidence for binarity exists for IRS13E2 and IRS13En.

But the most important factor was mentioned in Sect.\ \ref{res_WC}:
dust contamination. There is evidence of a large dust content in
IRS13E which may affect the infrared spectra of its stellar
components. As a consequence, the luminosities are most likely upper
limits. However, the luminosity difference between IRS13E4 and IRS7W
is 0.7 dex, or a factor of 5. Assuming that both stars have the same
luminosity and same spectral energy distribution, this means that the
total K band flux is 5 times larger than the stellar+wind flux. This
is exactly what is found by \citet{paulwc9} in their analysis of the
WC9 star HD164270: dust contributes 80\%\ of the continuum. Note that
their estimate is at 3 \mum\ and not in the K-band, where dust
contamination is smaller. But \citet{paulwc9} also claim that HD164270
has a weak dust shell compared to other dusty WC stars. Hence, for a
typical WC star a 80 \% dust contribution to the K band continuum is
not unrealistic.

Although further high spatial resolution photometric data in the
near/mid IR range are needed to establish the true SED of each
component, it is likely that dust explains most of the peculiar
properties of the IRS13E Wolf-Rayet stars.

\section{Metallicity in the central parsec}
\label{dis_z}

We first present a determination of the global metallicity in the GC
by means of WN stars, and we then discuss its effect on \hei.

\subsection{Metallicity from N content of WN stars}
\label{dis_z_N}

\citet{paco04} presented a new method for deriving metallicity in
evolved massive stars. Their idea relies on the fact that the surface
Nitrogen content of massive stars reaches a maximum during evolution
due to production of N through the CNO cycle. Observationally, this
maximum is obtained in WN stars. Its exact value is independent
of the initial mass and of the wind properties. It changes with the
initial metal content since the amount of Nitrogen synthesized is
directly linked to the initial CNO content. Hence, the comparison of
derived N abundances in WN stars to evolutionary tracks is an indirect
tracer of the initial metallicity.

In Table \ref{tab_res} we list our derived Nitrogen mass
fraction. Only for WN stars such an estimate could be performed due to
the presence of \ion{N}{iii} lines in the K band spectrum, especially
\niii. For most stars, we have $X(N) \approx 0.0135$, with the
exception of IRS15SW and AFNW which show a larger N content (up to
0.0326). However, for the latter star the uncertainty on the N
abundance determination is quite large due to the low resolution and
low S/N ratio of our spectrum. A direct comparison of this range of
values to predictions of surface enrichment in evolutionary models
\citep[see e.g. Fig.\ 4 of][]{paco04} indicates an initial metallicity
between solar and twice solar. If IRS15SW and AFNW are not taken into
account, then a solar metallicity is favoured. Strictly speaking, we
derive a lower limit since one cannot be sure that all WN stars have
reached the phase in which their surface N abundance is
maximum. Indeed, although this phase is long for initially very
massive stars, it is quite short for stars with M $\sim$ 20-50 \msun\
which is more appropriate for our WN8 stars. In conclusion, we
estimate the stellar GC metallicity to be \textit{at least}
solar. Stronger constraints cannot be derived from the present set of
data/models.

Stellar studies usually indicate a solar metallicity for the
GC. \citet{paco04} found $Z=Z_{\sun}$ for WN stars in the Arches
cluster. This is also comparable to the determinations of
\citet{carr00} and \citet{ramirez00} who also derived a solar [Fe/H]
from the study of Iron lines in red supergiants in the central 2.5 pc.

Abundances have also been derived from interstellar gas
studies. \citet{sf94} derived solar Ne content and twice solar Ar and
N based on photoionisation models aimed at reproducing nebular fine
structure lines in the mid infrared. However, improvements in the
knowledge of the Ar collisional strengths led to a revision of the Ar
abundance down to a nearly solar value \citep[see discussion
in][]{carr00}.  This is to be contrasted by the recent analysis of ISO
data by \citet{giveon02} who found Ne $\sim$ 1.4 Ne$_{\odot}$ and Ar
$\sim$ 2.5 Ar$_{\odot}$. Similarly, \citet{leticia03} revised the
Galactic metallicity gradients. Extrapolating their results to the
Galactic Center, one should expect $n(\rm{Ne/H}) \sim n(\rm{Ar/H})
\sim 1.5-2.5 Z_{\odot}$ and $n(\rm{N/H}) \sim 5 Z_{\odot}$.

Finally, observations of K shell Fe lines by Chandra have been used by
\citet{maeda02} to derive an Iron constant as large as 4 times solar
for the H~{\sc ii} region SgrA East. Such a large metallicity was
recently confirmed on a larger scale by X-ray observations with Suzaku
\citep{koyama06}.

Clearly, the measurement of the metallicity in the Galactic Center
needs more investigation. We have seen in Sect.\ \ref{16c} and
\ref{15ne} that individual elemental abundances could be super-solar
(Mg, Si). But the uncertainties associated to these estimates did not
allow a reliable statement. The use of high signal to noise, high
spectral resolution data is mandatory in order to resolve weak
metallic lines of Ofpe/WN9 stars, red supergiants as well as AGB
stars. The distinction between $\alpha$ element and Iron abundances is
also crucial. $\alpha$ elements are essentially produced in massive
stars while Iron is mainly synthesized in low mass stars during the
type Ia supernova phase. The ratio of $\alpha$ elements to Iron
abundances is thus an indirect tracer of the slope of the IMF.

\subsection{Metallicity effect on \hei}
\label{dis_z_hei}

\begin{figure}
\includegraphics[width=9cm]{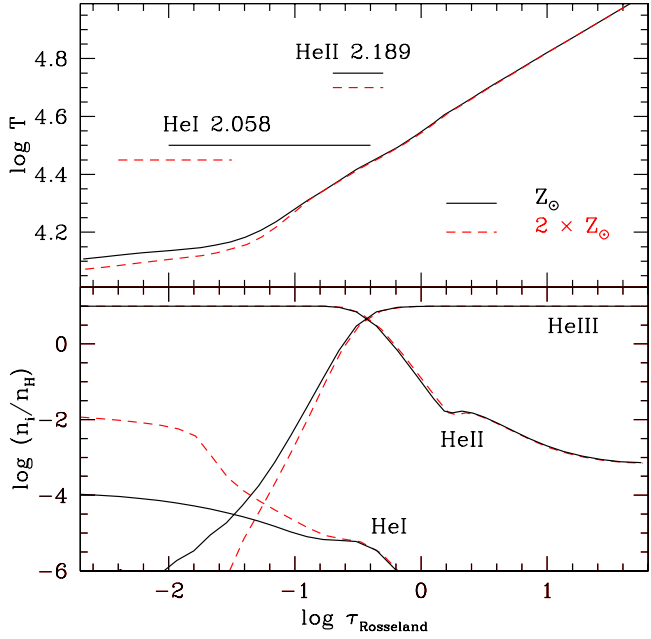}
\caption{Temperature (upper panel) and He ionisation (lower panel) structure of the best fit model for IRS9W. The solid line is the model with solar metallicity, while the dashed line is for $Z = 2 \times\ Z_{\odot}$. In the upper panel, the line formation regions of \hei\ and \heii\ are indicated by horizontal lines. \label{line_form_9w}}
\end{figure}

We have seen in Sect.\ \ref{9w} that \hei\ is extremely sensitive to
the metal content. Indeed, the higher the metallicity, the stronger
the line emission so that $Z = 2 \times\ Z_{\odot}$ was actually
preferred for IRS9W. Such a behavior was previously noted by
\citet{cbp98} and \citet{bc99} and was attributed to the sensitivity
of \hei\ to the EUV radiation field, which in turn depends critically
on the heavy-metal content. Fig.\ \ref{line_form_9w} shows the He
ionisation and temperature structure of the models for IRS9W with $Z =
Z_{\odot}$ and $Z = 2 \times\ Z_{\odot}$. This figure helps us
understand in detail what happens when the metal content is
increased. Due to the larger blocking of radiation by metal
opacities, the EUV flux is reduced which reduces the ionisation in the
outer part of the atmosphere (see lower panel of Fig.\
\ref{line_form_9w}). In addition, the escape of radiation through
metallic lines is favored so that the efficiency of line cooling is
enhanced, leading to a reduction of the temperature for $\log
\tau_{\rm Rosseland} \leq\ -1.0$ (upper panel of Fig.\
\ref{line_form_9w}). Both effects (line cooling and line blocking)
increase the \ion{He}{i} content in the outer atmosphere. In Fig.\
\ref{line_form_9w}, we also indicate the line formation region of
\hei\ and \heii. We see that \heii\ emerges from a zone where the
atmospheric structure is barely modified, so that its strength is
similar in both models (see Fig.\ \ref{fit_9w_K}). On the other hand,
\hei\ is produced in the outer atmosphere where the amount of He~{\sc
i} is larger in the $Z = 2 \times\ Z_{\odot}$ model. Consequently,
\hei\ is stronger in the high metallicity model. Note that this
explanation is fully consistent with the approach of \citet{paco94}:
when metallicity is increased, the \ion{He}{i} content increases which
leads to a decrease of the escape probability of the \ion{He}{i} 584
\AA\ line controlling the upper level of \hei\ -- \ion{He}{ii} being the
dominant ionisation state. This implies a stronger
emission at 2.058 \mum \citep[see Eq.\ 2 of][]{paco94}.

Does that mean that a twice solar metallicity should be adopted for
IRS9W? This would be an over-interpretation of the results. Indeed,
what the previous exercise reveals is that \hei\ is extremely
sensitive to the EUV radiation. We cannot be absolutely sure that we
correctly predict this part of the spectrum. Indeed, our models are
limited in the sense that we cannot include \textit{all} lines from
\textit{all} elements. \hei\ may also suffer from inaccuracies in
metal atomic data. Hence, it may well be that increasing the metal
content artificially compensates for the lack of metallic lines /
elements. Besides, we have discussed only one example for which a
larger metal content improves the fit of \hei. But there are other
stars (e.g. IRS7E2) for which having a higher Z does not help, \hei\
remaining too weak. Hence, we conclude that the metallicity cannot be
accurately derived from its effect on \hei.

\section{Wind properties of Wolf-Rayet stars}
\label{dis_wind}

We have derived the wind properties of a homogeneous and reasonably
large sample of Wolf-Rayet stars, mainly composed of WN and Ofpe/WN9
stars. In order to see if the GC stellar winds are similar to other
Galactic WR stars, we inspect the behavior of the terminal velocities
and mass loss rates (corrected for clumping).

\begin{figure}
\includegraphics[width=9cm]{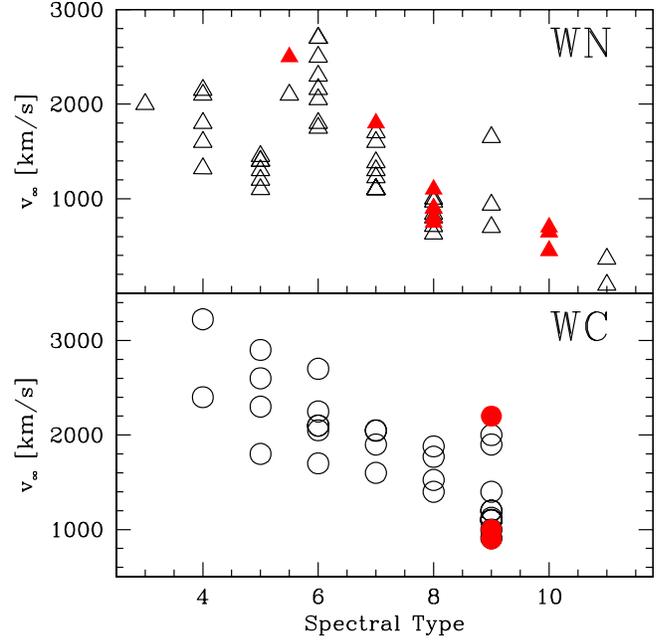}
\caption{Terminal velocity of GC WR stars (filled symbols) and other Galactic stars (open symbols) as a function of spectral type. The data for galactic stars are taken from the compilation of \citet{vdh}. Upper panel: WN stars; Lower panel: WC stars.\label{vinf_wr}}
\end{figure}

Terminal velocities for other Galactic stars are taken from the
catalog of \citet{vdh} and are compared to our sample in Fig.\
\ref{vinf_wr}. We see that there is a remarkable agreement. We have
seen in Sect.\ \ref{dis_ofpe} that \vinf\ is usually larger in the GC
than in the LMC for Ofpe/WN9 stars, but this was at least
qualitatively explained by a different metallicity/evolutionary
state. We then conclude that the terminal velocities of the GC
Wolf-Rayet stars are identical to other Galactic stars.

\begin{figure}
\includegraphics[width=9cm]{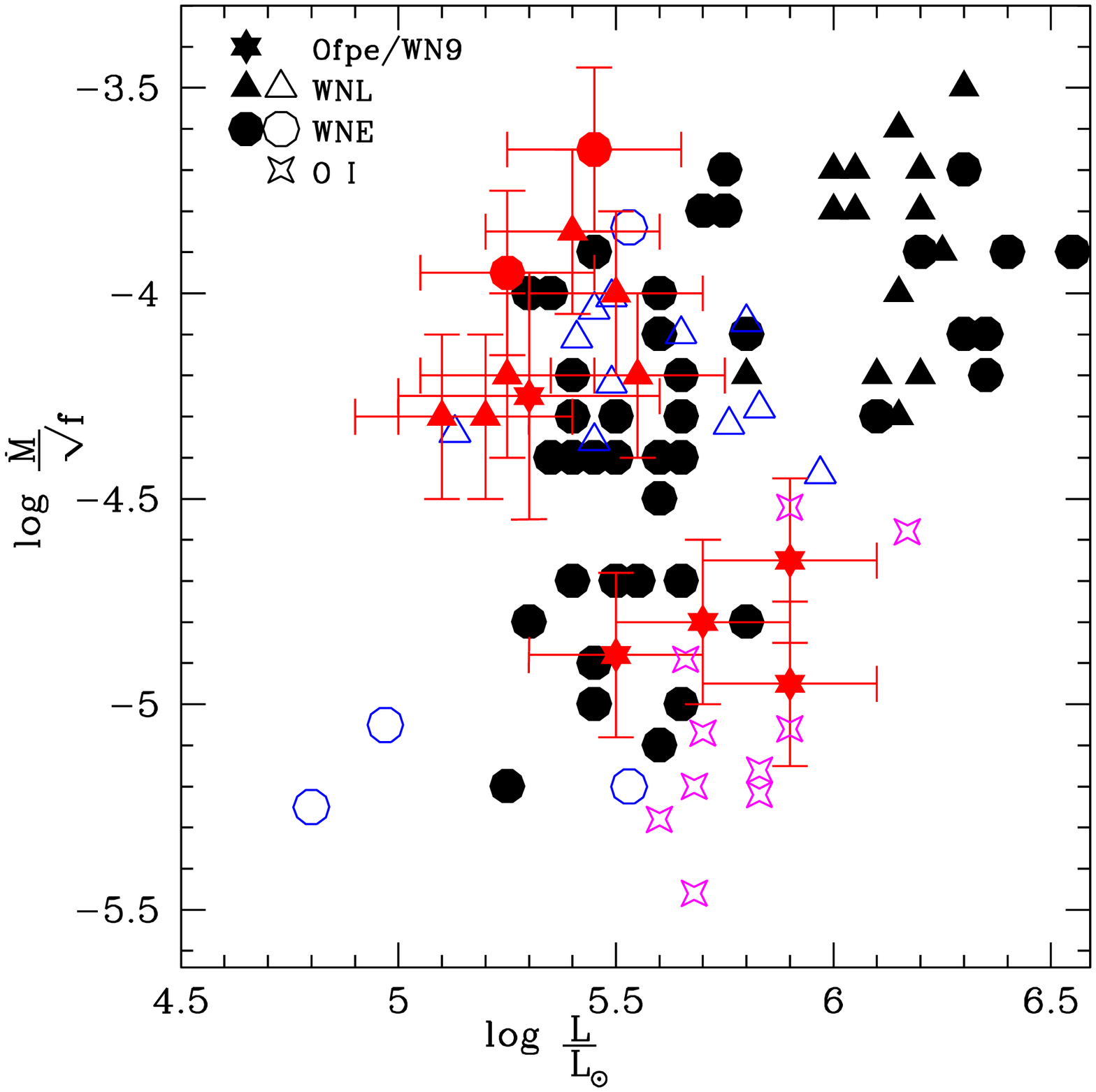}
\caption{Mass loss rate corrected for clumping ($\dot{M}/\sqrt{f}$) as a function of luminosity for GC WN stars (red filled symbols) compared to other Galactic Wolf-Rayet stars (blue and black symbols) and O supergiants (magenta star symbols). Comparison data from \citet{paul95_wnl}, \citet{paul95_wne}, \citet{morris00} for WN stars and \citet{repolust04}, \citet{jc05} for O supergiants are shown by open symbols. The results of \citet{hamann06} (WN stars) are displayed by the black filled symbols. Early WN stars ($\leq$WN7) are shown by triangles, while late WN stars ($\geq$WN8) are circles. Ofpe/WN9 stars are star symbols and O supergiants are open asterisks.  \label{mdot_L}}
\end{figure}

The mass loss rates of WN and Ofpe/WN9 stars are shown in Fig.\
\ref{mdot_L} as a function of luminosity \footnote{our sample of WC
stars is too small for a comparison to other Galactic stars to be
relevant}. We also plot the results of \citet{paul95_wne},
\citet{paul95_wnl}, \citet{morris00}, (all in blue open symbols) and
\citet{hamann06} (black filled symbols). The GC mass loss
rates are in good agreement with the Crowther et al./Morris et al.\
results for late WN stars (triangles). For early WN stars, the samples
are too small to draw any conclusion. On the contrary,
there is a large discrepancy between our results and those of
\citet{hamann06}, at least for late WN stars. This difference is not so
much due to the mass loss rates themselves as it is to
luminosities. Indeed, both samples have similar values of
$M_{\odot}/\sqrt{f}$, but the luminosities in the GC are lower. Since
most of the stars of the Hamann et al.\ sample have poorly constrained
distances (only a few are in clusters), while the distance to the GC
is accurately known, we argue that the main reason for the observed
difference is an overestimate of the luminosities in the Hamann et
al.\ sample. An additional explanation to the GC/Hamann et al.\ WNL
luminosity difference could be that their sample contains several
H-rich WN stars. These stars are thought to be very massive stars with
a Wolf-Rayet appearance but possibly still core H burning objects
on/close to the main sequence \citep[e.g.][]{moffat06}. Such stars are
expected to be very luminous. However, they usually have spectral
subtypes earlier than WN7 \citep[e.g.][]{cd98}. A significant
fraction of the very luminous stars in the Hamann et al.\ sample
(represented by filled triangles around \lL\ $\sim$ 6.0-6.2 in Fig.\
\ref{mdot_L}) do not fall into this category (they are later WN
stars). Their luminosity is thus most likely overestimated due to
poorly known distance.

In Fig.\ \ref{mdot_L}, we also plot the mass loss rates of the GC
Ofpe/WN9 stars together with Galactic O supergiants. We see that in
terms of \mdot, the Ofpe/WN9 stars are closer to O stars than to WN
stars. This is another indication that Ofpe/WN9 stars are precursors
of WN8 stars as we have argued in Sect.\ \ref{dis_wn8}.

\section{He abundance in OB stars}
\label{dis_ob}

We have seen in Sect.\ \ref{OBIa} that the stellar and wind parameters
of the average OB stars in the central parsec were not strongly
constrained. Most of them have only upper limits (\teff, \mdot). For
the helium content, we could estimate $X(He) \simeq 0.2-0.35$. This
value is larger than the standard solar abundance (0.1).

Since OB supergiants are massive stars evolving away from the main
sequence, chemical enrichment is the most likely explanation. The
question is whether or not the amount of He observed is consistent
with such a mechanism. In that respect, the inclusion of rotation in
evolutionary models is a key ingredient since it triggers additional
mixing which modifies the He surface abundance X(He). In the recent
models of \citet{mm05}, X(He) evolves slowly from the initial value
during all the ``O'' phase, and then dramatically increases when the
star enters the Wolf-Rayet phase. At the end of the O phase, a He
content as large as 0.3-0.4 can be obtained. This is compatible with
our derived value. We have also seen in Sect.\ \ref{OBIa} that the
average projected rotational velocity of the brightest OB supergiants
was of the order 100 \kms. Inspection of Fig.\ 1 of \citet{mm03}
reveals that after 4-8 Myrs (age of the GC population of early type
stars), stars with initial masses in the range 25-60 \msun\
(appropriate for the OB supergiants) have indeed \vsini\ between 40
and 200 \kms\ (depending on the exact age / mass). This is another
indication that 1) the evolutionary tracks with rotation are adequate
for the present discussion, and 2) that the GC massive stars do not
have exceptionally large rotational velocities, as could be suspected
for such a dense environment where interactions should be frequent.

\begin{figure}
\includegraphics[width=9cm]{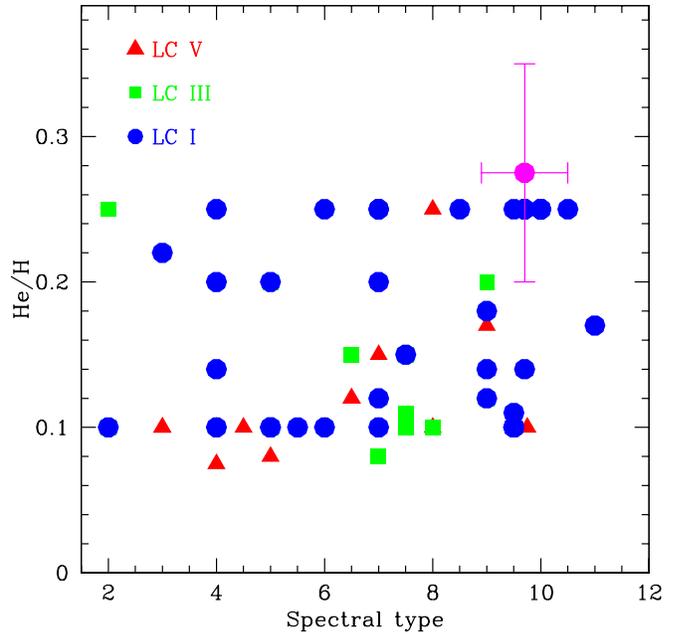}
\caption{Comparison between derived ratio of He to H abundance (symbol with error bars) of OB supergiants in the Galactic Center to values from other studies \citep{her01,jc03,evans04,markova04,repolust04}. Red triangles (green squares; blue circles) are dwarfs (giants; supergiants).\label{HeH_liter}}
\end{figure}

How does our He determination compare to other analyses of massive
stars? The answer is given in Fig.\ \ref{HeH_liter}. Blue circles show
the He content of Galactic O supergiants. It is important to note that
the uncertainty usually quoted in such determinations is
$\pm0.05-0.1$. The circle with the error bars in Fig.\ \ref{HeH_liter}
represents the average OB supergiants in the GC. We see that although
large, the He content of these stars is still compatible with the
range of values found for O supergiants. In conclusion, the GC OB
supergiants are \textit{on average} similar to other Galactic stars in
terms of He enrichment. This abundance is in addition compatible with
present evolutionary tracks.

\citet{pgm06} noted that a few OB supergiants had a strong He
absorption on the blue side of \brg\ and suggested that these stars
are significantly He-rich. Unfortunately, given the low S/N ratio
spectra of these stars, we could not derive quantitative constraints
on the He abundance.

\section{Conclusion}
\label{conc}

We have carried out a detailed analysis of 18 evolved massive stars
(Ofpe/WN9 and Wolf-Rayet) located in the central parsec of the Galaxy
from H and K band spectroscopy obtained with SINFONI on the
ESO/VLT. The average spectrum of 10 bright OB supergiants was also
examined. For quantitative analysis, we have used state of the art
non-LTE atmosphere models including winds and line-blanketing computed
with the code CMFGEN. The main results are:

\begin{itemize}

\item[$\bullet$] The population of massive stars in the central parsec
is able to supply the amount of H and He{\sc i} ionising
photons required to reproduce the nebular emission. The contribution
of the bright, cool post-main sequence massive stars, first discovered
by \citet{forrest87} and \citet{ahh90} and analysed by
\citet{paco97}, accounts for only $\sim$4\% of the total H ionising flux
reconciling stellar evolution with observations in the Galactic
Center. State of the art evolutionary and atmosphere models also
reconcile the observed stellar content with a population synthesis
model of a starburst of age $\sim$ 6 Myrs.

\item[$\bullet$] Ofpe/WN9 stars
are evolved massive stars close to a LBV phase. Compared to
\citet{paco97}, we find they are less He rich, slightly hotter
(although the temperature is poorly constrained) and have lower mass
loss rates. WN8 stars are found to have properties similar to other
Galactic Wolf-Rayet stars of the same spectral type. From
morphological as well as quantitative arguments, they are likely the
descendents of Ofpe/WN9 stars. Two stars are classified as WN8/WC9
stars. Their quantitative analysis reveals that, as suggested by their
spectral type, they show both H and He burning products at their
surface. A direct evolutionary link between the GC Ofpe/WN9, WN8 and
WN/C stars of the form
\begin{center}
(Ofpe/WN9 $\rightleftharpoons$ LBV) $\rightarrow$ WN8 $\rightarrow$ WN/C
\end{center}
is proposed, similar to \citet{paul95_wnl}.

\item[$\bullet$] Quantitatively, stellar evolutionary tracks with
rotation and $Z = Z_{\odot}$ overpredict the luminosity of the GC
Wolf-Rayet stars. Tracks with $Z = 2 \times\ Z_{\odot}$ are more
appropriate. This may indicate that the mass loss rates adopted in the
current evolutionary tracks during the Wolf-Rayet phase are too
low. However, accurate metallicity determinations are needed to solve
this issue.

\item[$\bullet$] On average, GC OB supergiants are He rich, but not
significantly richer than other galactic supergiants. Their present
projected rotational velocities is $\sim$ 100 \kms. These properties
are quantitatively compatible with stellar evolution with rotation on
and close to the main sequence.

\item[$\bullet$] The wind properties of WN stars in the Galactic
center are very similar to other Galactic stars of the same spectral
type. The luminosities of late WN stars are lower in the GC than in
the sample of \citet{hamann06} but in good agreement with the results
of \citet{paul95_wnl}. We argue that the difference with the results
of Hamann et al. is due to uncertain distances for their sample stars,
while in our case the distance to the GC is well constrained.

\end{itemize}

Our study has shown that the GC massive stars are, on average, similar
to other Galactic stars. This strongly suggests that they follow a
common evolution, regardless of their possible different formation
process. A key question which remains to be addressed in future
studies is the metallicity in the GC. A better knowledge of this
parameter is crucial to quantitatively test evolutionary models. It is
also crucial in the context of the chemical evolution of the
Galaxy. Also important is the study of nebular emission in the GC by
means of recently developed 3D photo-ionization models. Our total
stellar SED will be a crucial input for such models which will likely
constrain the geometry and density of the ionised gas.

\begin{acknowledgements}

We thank F. Najarro for interesting discussions. FM acknowledges
support from the Alexander von Humboldt foundation.

\end{acknowledgements}

\bibliography{biblio.bib}

\end{document}